\PassOptionsToPackage{table,xcdraw}{xcolor}
\documentclass[manuscript,screen]{acmart}
\AtBeginDocument{%
  \providecommand\BibTeX{{%
    \normalfont B\kern-0.5em{\scshape i\kern-0.25em b}\kern-0.8em\TeX}}}

\usepackage[document]{ragged2e}
\usepackage{fontawesome}
\usepackage{tikz}
\usepackage{array}
\usepackage{multirow}
\usepackage{multicol}
\usepackage{graphicx}
\usepackage{subcaption}
\usepackage{titlesec}
\usepackage[table]{xcolor}
\usepackage[many]{tcolorbox} 
\usepackage[flushleft]{threeparttable}
\newcommand{\faIcon}[1]{\faicon{#1}}
\usepackage[normalem]{ulem}
\useunder{\uline}{\ul}{}
\usepackage{lscape}
\usepackage{longtable}
\newcolumntype{P}[1]{>{\RaggedRight\hspace{0pt}}p{#1}}

\definecolor{tableGray}{RGB}{243, 244, 245}
\definecolor{borderGray}{RGB}{229, 230, 233}

\newtcolorbox{boxA}{
    colback = tableGray, 
    boxrule = 0pt  
}

\begin{document}


\title [Enablers and Barriers of Empathy in Software Developer-User Interactions]{Enablers and Barriers of Empathy in Software Developer and User Interactions: A Mixed Methods Case Study}

\author{Hashini Gunatilake}
\email{hashini.gunatilake@monash.edu}
\orcid{0000-0002-4823-0214}

\affiliation{%
  \institution{Faculty of Information Technology, Monash University}
  \country{Australia}
}

\author{John Grundy}
\orcid{0000-0003-4928-7076}
\affiliation{%
  \institution{Faculty of Information Technology, Monash University}
  \country{Australia}}
\email{john.grundy@monash.edu}

\author{Rashina Hoda}
\orcid{0000-0001-5147-8096}
\affiliation{%
  \institution{Faculty of Information Technology, Monash University}
  \country{Australia}
}
\email{rashina.hoda@monash.edu}

\author{Ingo Mueller}
\affiliation{%
  \institution{Faculty of Information Technology, Monash University}
  \country{Australia}}
\email{ingo.mueller@monash.edu}

\renewcommand{\shortauthors}{Gunatilake, et al.}


\begin{abstract}
Software engineering (SE) requires developers to collaborate with stakeholders, and understanding their emotions and perspectives is often vital. Empathy is a concept characterising a person's ability to understand and share the feelings of another. However, empathy continues to be an under-researched human aspect in SE. We studied how empathy is practised between developers and end users using a mixed methods case study. We used an empathy test, observations and interviews to collect data, and socio–technical grounded theory and descriptive statistics to analyse data. We identified the nature of awareness required to trigger empathy and enablers of empathy. We discovered barriers to empathy and a set of potential strategies to overcome these barriers. We report insights on emerging relationships and present a set of recommendations and potential future works on empathy and SE for software practitioners and SE researchers.

\end{abstract}




\maketitle

\section{Introduction} \label{SEC:Introduction}

Empathy has been recognised as a human aspect that can help to understand software developer and stakeholder human interactions \cite{gunatilake2023empathy}. It is also considered as a vital competency in many professional fields such as medicine \cite{hojat2018jefferson, medical1999learning}, healthcare \cite{nunes2011healthdisciplines}, nursing \cite{yu2009evaluation}, animal science \cite{hazel2011animalscience}, education \cite{blanco2017deconstructing, levy2018educating}, professional writing and reviewing \cite{de2007professional}, marketing \cite{delpechitre2013review}, and project management \cite{ewin2021empathy}. Empathy has been recognised as factor that improves inter-group attitudes and relationships \cite{batson2009intergroup}. 

Despite its prominence across disciplines, empathy remains a complex phenomenon with no one unified definition \cite{clark2019feel, cuff2016empathy, neumann2015measures, preston2002empathy, decety2007empathic, guthridge2021taxonomy}. One definition of empathy is \textit{``the ability to experience the affective and cognitive states of another person, while maintaining a distinct self, in order to understand the other''} \cite{guthridge2021taxonomy}. This meta definition is derived from an inductive conceptual content analysis of the existing definitions, and provides a clearer outline of its fundamental dimensions.


There are only a limited number of studies to date on empathy in SE \cite{gunatilake2023empathy}. Empathy has been discussed in the context of user experience (UX) with the use of personas \cite{karolita2023personas, ferreira2015designing}, used in combination with empathy map to improve personas development \cite{ferreira2016pathy}, as a tool for design thinking (DT) to better understand the requirements, thereby enhancing software quality \cite{canedo2020design} and addressing user privacy concerns \cite{levy2018importance}. In another study, researchers investigated the relationship between the collective empathy of software development teams and the effectiveness of their project process \cite{akgun2015collectiveempathy}. Here, collective empathy encompasses cognitive, affective, and behavioural empathy. They found that collective empathy impacts their project management effectiveness including team learning, product speed-to-market, and reduced project development costs. 



A common thread across these works is that \textit{empathy seems vital for fostering human connections} \cite{hojat2016empathy} and that \textit{having a good connection among software project stakeholders} may positively influence the success of a project. However, to the best of our knowledge, there are no studies that study empathy in the critical relationship of \textit{software developers and end users}, exploring its enablers and barriers \cite{gunatilake2023empathy}. Due to the limited research in SE and the positive impact of empathy observed in other disciplines, we were motivated to explore this topic. To do this we conducted a mixed methods case study by exploring how empathy is practised between developers and end users in a specific software development project. Our study centred around the following research questions: 

[RQ] To what extent do developers demonstrate empathy towards end users, and conversely, how empathetic are end users towards developers in the context of software development?
\begin{itemize}
    \item What factors contribute to the empathy between software developers and end users?
    \item What factors impede empathy between software developers and end users?
    \item What steps were taken to address the factors that impede empathy between developers and end users?
\end{itemize}

In the absence of a universally accepted definition, we used the meta definition of empathy described earlier \cite{guthridge2021taxonomy}. We employed the questionnaire of cognitive and affective empathy (QCAE), conducted observation of empathy cues \cite{elliott2011empathy, nicolai2007rating, chartrand1999chameleon}, and interviews with developers and end users to collect both qualitative and quantitative data. We used socio–technical grounded theory (STGT) for qualitative data analysis \cite{hoda2021STGT} to analyse the qualitative data and measures of dispersion in descriptive statistics for our quantitative data analysis. 

The key contributions of this work include the identification of: 
\begin{itemize}
    \item The awareness required to trigger empathy and enablers of empathy between developers and end users;
    \item Some key potential barriers to empathy of developers and end users, and strategies to overcome these barriers;
    \item A set of actions to manage empathy enablers and barriers in SE projects; and 
    \item A set of recommendations derived from our findings for software practitioners and SE researchers.
\end{itemize}

The rest of this paper is organised as follows. Section \ref{SEC:Definitions} introduces definitions of key empathy specific terminology used throughout this paper. Section \ref{SEC:Related Works} presents an overview of key related studies. Section \ref{SEC:Research Design} describes our research methodology. Section \ref{SEC:Findings} describes the key findings from this work, and Section \ref{SEC:Discussion} discusses our insights,  implications, recommendations for practitioners \& researchers and directions for future work. In Section \ref{SEC:Threats to Validity}, we present limitations of this study, followed by our conclusions in Section \ref{SEC:Conclusion}.

\section{Glossary of Terms} \label{SEC:Definitions}

We use a range of empathy-specific terminology throughout our paper and brief definitions of these terms are provided in Table \ref{TAB:Glossary of Terms}. 

\begin{table*}[ht]
\begin{threeparttable}
    \footnotesize
    \centering
    \caption{Glossary of Terms}
    \label{TAB:Glossary of Terms}
    \begin{tabular}{ll}
        \toprule
         \textbf{Term} & \textbf{Definition}  \\
         \midrule
         
         Empathy & The ability to experience affective and cognitive states of another person, while maintaining a distinct self, in order to\\
         & understand the other \cite{guthridge2021taxonomy}.\\
         
         Cognitive Empathy & The ability of a person to consciously detect and understand the internal states of others \cite{goldman2011two, wallmark2018neurophysiological}.\\

         Affective/Emotional Emp. & The ability of a person to perceive and share other individual’s emotional states and feelings \cite{de2008putting, ilgunaite2017measuring}.\\

         Behavioural Empathy & This consists of two types of empathic behaviours i.e., \emph{behaviour mirroring} and \emph{empathic communication}. Mimicking of facial \\
         & expressions, mannerisms, postures, \& gestures of other person is referred as behaviour mirroring. Empathic communication\\
         & is defined as intentional behaviour that displays cognitive and/or affective empathy towards the other person \cite{clark2019feel}.\\

         Sympathy & An emotional response stemming from the apprehension of another’s emotional state or condition that is not the same as the\\ 
         & other’s state or condition, but consists of feelings of sorrow or concern for the other \cite{eisenberg2009empathic, eisenberg2014sympathy}.\\

         Perspective Taking\textsc{*} & The ability of a person to see the situation from another person’s perspective \cite{reniers2011QCAE, ilgunaite2017measuring}.\\

         Online Simulation\textsc{*} & The ability of a person to understand and mentally represent how another person is feeling \cite{reniers2011QCAE, ilgunaite2017measuring}.\\

         Emotion Contagion\textsc{*} & The ability of a person to reflect self oriented emotions while noting the emotional states of others \cite{reniers2011QCAE, ilgunaite2017measuring}.\\

         Proximal Responsivity\textsc{*} & A person’s emotional reaction to the moods of another person, who is physically or emotionally close to this person \cite{reniers2011QCAE, ilgunaite2017measuring}.\\
    
        Peripheral Responsivity\textsc{*} & A person's emotional reaction to the state of moods of another person, who is not close to them or they do not know that \\
        & person at all \cite{reniers2011QCAE, ilgunaite2017measuring}.\\
    
        Empathic Concern & The tendency of a person to experience feelings of warmth, compassion and concern for others undergoing negative\\
        & experiences \cite{davis1980IRI}.\\
    
        Emotional Intelligence & The ability to monitor one's own and others' feelings and emotions, to discriminate among them and to use this information\\
        & to guide one's thinking and actions \cite{salovey1990emotional}.\\
    
        Social Awareness & The awareness of others’ emotions \cite{goleman2002emotional}.\\
    
        Collectivism & ``Societies in which people from birth onwards are integrated into strong, cohesive in-groups, which throughout people’s \\
        & lifetime continue to protect them in exchange for unquestioning loyalty'' \cite{hofstede1991empirical}.\\
    
        Individualism & ``Societies in which the ties between individuals are loose; everyone is expected to look after themself \& their immediate\\
        & family'' \cite{hofstede1991empirical}.\\

         \bottomrule
    \end{tabular}
    \begin{tablenotes}
      \item \textit{\textsc{*} A subscale of QCAE}
    \end{tablenotes}
    \end{threeparttable}
\end{table*}

\section{Related Work} \label{SEC:Related Works}
Empathy is widely regarded as a multidimensional construct \cite{clark2019feel, cuff2016empathy}, with \textit{cognitive}, \textit{affective} and \textit{behavioural} empathy as its key three dimensions \citep{clark2019feel}. Others have argued there are only two dimensions to empathy, cognitive and affective \cite{cuff2016empathy}. There are two key commonly used methods to measure empathy: measuring empathy via self-assessments, and via neurophysiological examination methods of studying different brain activities using brain images. Researchers have also measured empathy through observations, based on demonstrated empathy behaviours. These empathy behaviours can be divided into two main categories, verbal and nonverbal behaviours, that have been widely studied \cite{elliott2011empathy, nicolai2007rating, chartrand1999chameleon, decety2004functional}. 

Verbal empathy can be broken down into four behaviours \cite{elliott2011empathy}, empathic understanding responses, empathic affirmations, empathic evocations, and empathic conjectures. Non-verbal empathy behaviours consider \textit{sounds and signals} to understand whether people show interest in others \cite{nicolai2007rating}. In their study, attentive listening was identified as one such a behaviour including: sounds as verbal acknowledgements without interrupting the flow of other person's speech; repeated nodding, tilted head, and placing thumb and forefinger of one hand on the chin as signals for attentive listening \cite{nicolai2007rating}. Wide open eyes and raised eyebrows also have been classified as signals for attentive listening \cite{ekman1971constants} and for people feeling addressed and therefore actively participating \cite{bartel2000collective}. These mimicking behaviours are known to occur in combination with each other and often associated with an activated posture such as bending forward towards the other person. In addition to the sounds and signals of attentive listening, Chartrand et al., explained \textit{similar facial expression, posture or gesture} as a nonverbal empathy behaviour \cite{chartrand1999chameleon}.

Research across disciplines supports strong association between \textit{similarities} and empathic behaviour, such as in psychology \cite{batson1996prior, eklund2009similar, heinke2009cultural, hoffman2001empathy, yaghoubi2023young}, neuroscience \cite{preis2012pain, xu2009pain, yaghoubi2021histories}, and philosophy \cite{snow2000empathy}. Similarities can be defined as sharing a similar experience or demographic characteristic with another person \cite{jami2023interaction}. The study by Motomura et al. highlighted the nuanced effects of familiarity on empathy valence, distinguishing between positive and negative empathy \cite{motomura2015interaction}. Specifically, it found that positive empathy, linked to activities like nursing or support, is less likely to occur with strangers. In contrast, negative empathy, associated with adverse situations such as danger, can readily happen even with unfamiliar individuals. Investigating empathic brain activity, the study suggested that \textit{familiarity} acts as an enabler for positive empathy. Building on this, researchers have argued that the concept of \textit{relatedness} is crucial for human empathy, going so far as to propose that familiarity can be more accurately defined as relatedness \cite{airenti2015cognitive}.


The studies on \textit{culture} and empathy reported that cultural backgrounds of people influence their empathic behaviour \cite{jami2023interaction}. It is commonly believed that empathic behaviours are more valued in collectivistic societies than individualistic societies \cite{halabi2008social, kitayama2000culture, schwartz2010biculturalism}. A study found a positive relationship between affective empathy and country-level index of collectivism, but no association was found between cognitive empathy and country-level index of collectivism/individualism \cite{chopik2017differences}. However, another study found that individualistic societies have higher empathy compared to collectivistic societies \cite{cassels2010role}. This was due to their ability to have an independent cognitive and emotional state by differentiating their own emotions from others. However, most of the studies on culture and empathy have found higher associations between collectivism and dispositional affective and cognitive empathy \cite{chopik2017differences, wu2007effect}. 

Empathy barriers have been studied primarily in healthcare domain \cite{derksen2016managing, howick2017barriers, halpern2003clinical}. A study on healthcare identified \textit{time pressure, conflicting priorities, bureaucracy} as barriers to empathy \cite{howick2017barriers}. \textit{Anxiety}, \textit{inability to understand the relationship between illness and patients' emotional needs}, and \textit{negative emotions due to tensions with patients} have been identified as barriers to physicians' empathy towards patients \cite{halpern2003clinical}. This study found that physicians become anxious due to time pressure hence there is a tendency of not listening to patients. Another study found empathy barriers of oncology nurses. This study found three types of barriers: those related to nursing, healthcare, and cancer care \cite{taleghani2018barriers}. 
\textit{Lacking compassion, disinterest in oncology nursing \& self-criticism, psychological distress} were identified as barriers related to nursing. Barriers related to healthcare included \textit{job strain, task centeredness rather than patient-centeredness, lack of formal training for empathy with cancer patients, lack of manager support, and nurse-patient gender imbalance}. Barriers related to cancer care were \textit{difficulty of maintaining empathy with cancer patients and feeling of uselessness of care for cancer patients}.
In summary, familiarity, similarities, relatedness, culture were identified as empathy enablers, and time pressure, negative emotions, disinterest, task centredness were identified as empathy barriers. 

Many empathy measures, models and techniques have been developed over the years. In our previous work, we developed a preliminary empathy taxonomy for SE considering these empathy models, techniques and measures \cite{gunatilake2023empathy}. Since empathy is seen as beneficial for improving human connections across disciplines, and has not been adequately studied in SE, we became interested to explore it in the context of one of the prominent relationships in SE contexts, between software developers and end users.




\section{Research Design} \label{SEC:Research Design}
We wanted to understand how empathy is practised between software developers and end users. Our topic lend itself well to a mixed methods case study, where we could investigate the phenomenon in depth, from multiple angles using multiple approaches to understand it up-close and in depth. 
We adhered to the guidelines outlined by Runeson et al. \cite{runeson2009guidelines} when conducting this case study. 


\subsection{Case Description} \label{SEC:Case Description}

\textit{The Project:} The selection of the \textit{AskPCOS project} aligns with our research questions on factors influencing empathy between software developers and end users. Access to end users is notoriously challenging to arrange and often limits what can be researched and how. Our access to the AskPCOS project provided us the level of access required to do this in a real-world SE case setting. This project's context offers a practical environment to explore the dynamics of empathy, ensuring our study is grounded in real-world complexities and user engagement constraints. Thus, the AskPCOS project serves as an ideal and relevant case, allowing us to delve into the nuanced factors that contribute to or impede empathy between software developers and end users.
The project involved the development of an engaging extension that integrates SMART (Specific, Measurable, Achievable, Relevant, and Time-Bound) goal setting process into the existing AskPCOS web application.\footnote{https://www.askpcos.org/} This solution helps women with Polycystic Ovary Syndrome (PCOS) to better manage their lifestyle. This software project intended to digitise questionnaires, checklists, and other paper based material related to PCOS lifestyle education sessions conducted by the Monash Health public PCOS clinic. 

The fourth author supervised this software project, facilitating our access to it. Healthcare professionals from Monash Centre for Health Research and Implementation (MCHRI)\footnote{https://www.monash.edu/medicine/mchri} advertised the project in Facebook groups dedicated to PCOS patients. The PCOS patients who expressed interest were then recruited as study participants. Throughout the study, we maintained a consistent group of PCOS patients.

The development project was carried out over a period of 24 weeks and there were two iterations within it. The main stakeholders in the development project were Healthcare professionals from MCHRI (domain experts) and two academic staff from Monash Art, Design \& Architecture (MADA) faculty.\footnote{https://www.monash.edu/mada} Our study had several multi-disciplinary participants: two Masters level design students from MADA (UX designers), six final year undergraduate IT students from Monash Faculty of Information Technology (FIT) (developers), and four patients with PCOS (target end users). These stakeholders and participants had multiple interactions throughout the project. 

\textit{The People:} Our study\footnote{Approved by Monash Human Research Ethics Committee. ERM Reference Number: 32281} was based on the usability testing sessions conducted with the participation of software developers and potential end users. The purpose of these sessions were to evaluate the usability of the newly implemented SMART goal setting process of the AskPCOS web application. Usability testing sessions provided a platform for developers and users to directly interact with each other. All developers had the opportunity to directly interact with at least one user by acting as the hosting developer of the usability testing sessions. Likewise all users were able to interact with developers. All participants stated that they were able to build a better connection due to interaction and/or familiarity with each other. 

\begin{table*} [htbp]
    \caption{Demographics of the Interview Participants}
    \label{TAB:Demographics of the Interview Participants}
    \footnotesize
    \begin{tabular}{llllllllll}
        \toprule
        \textbf{Participant} & \textbf{Age} & \textbf{Education} & \textbf{Experience} & \textbf{Ex. in} &  \textbf{Ex. in} & \textbf{Ex. in} & \textbf{Ex. in} & \textbf{Affinity} & \textbf{Familiar} \\
        
        \textbf{ID} & \textbf{Group} & & \textbf{in working} & \textbf{working} &  \textbf{working} & \textbf{working} & \textbf{using} & \textbf{to tech.} & \textbf{with}\\
       
        & & & \textbf{with teams} & \textbf{with people} &  \textbf{with patients/} & \textbf{with} & \textbf{eHealth} & \textbf{vs} & \textbf{software} \\

        & & & & \textbf{from diverse} &  \textbf{people with} & \textbf{customers/} & \textbf{apps} & \textbf{people} & \textbf{developers}\\

        & & & & \textbf{backgrounds} &  \textbf{sensitive cond.} & \textbf{users} & & & \\
        
        \midrule
        \textbf{\textit{Developers}}  & &  &  &  &  &  & &  \\
        \midrule
        
        P1 & 20-25 & SD\textsc{*}, BE\textsc{*} & Yes & Yes & No & No & No & SHC\textsc{*} & -\\
    
        P2 & 20-25 & SD, BE & Yes & Yes & No & No & Yes & STC& -\\

        P3 & 20-25 & DD\textsc{*}, BE+BCom\textsc{*} & Yes & Yes & No & No & No & STC & -\\

        P4 & 20-25 & SD, BE & Yes & Yes & Yes & Yes & No & STC& -\\

        P5 & 20-25 & DD, BE+BCom & Yes & Yes & No & Yes & No & STC & -\\

        P6 & 20-25 & DD, BE+BCom & Yes & Yes & No & No & No & FTC & -\\

        \midrule
        \textbf{\textit{Users}}   & &  &  &  &  &  & & \\
        \midrule
        PU1 & 20-25 & - & Yes & Yes & - & - & Yes & SHC & No\\

        PU2 & 31-35 & - & Yes & Yes & - & - & Yes & FTC & Yes\\

        PU3 & 41-45 & - & Yes & Yes & - & - & Yes & FHC & No\\

        PU4 & 26-30 & - & Yes & Yes & - & - & No & FHC & No\\
       
        \bottomrule
    \end{tabular}
    
     \begin{flushleft}
     \textit{\textsc{*}SD: Single Degree, DD: Double Degree, BE: Bachelor of Engineering, BCom: Bachelor of Commerce, FHC: Fully Human Centric, SHC: Somewhat Human Centric, STC: Somewhat Technology Centric, FTC: Fully Technology Centric}
    \end{flushleft}
    
\end{table*}
 
Table \ref{TAB:Demographics of the Interview Participants} shows key participant information. We interviewed six developers who were final year SE students and four end users who were females with the PCOS condition. A majority of developers and end users were from Australia except for one developer from Hong Kong and a user from Zambia. All the developers and end users were residing in Australasia while conducting this study. All the developers belonged to the age group 20-25 and there was only one female developer among them. End users were from four different age groups including 20-25, 26-30, 31-35 and 41-45. Half of the developers were following a single degree which is Bachelor of Engineering (BE) in SE and rest of the developers were following double degrees. They were following Bachelor of Commerce majoring in different areas like business analytics, finance and econometrics along with the Bachelor of SE. All the developers and users confirmed that they had prior experience in working in a team and with people from diverse backgrounds. Only one developer had prior experience in working with patients or other groups of people who tend to have some sensitive condition. Two developers had prior experience in working with customers/users in other university classes or in industry experience. Except for one user, the rest of the users had not interacted with software developers in any form prior to the study. Only one developer had used an eHealth app before, however three out of four end users had prior experience in using eHealth apps. From all the participants we inquired the affinity to technology and people on a scale of zero to three, where zero was fully human-centric, one was somewhat human centric, two was somewhat technology-centric and three was fully technology-centric. Most of the developers self-identified as technology-centric and most of the users self-identified as human-centric.

\textit{The Process:} 
In the first 12 weeks, design students from MADA had discussions with healthcare professionals from MCHRI and end users to get a better understanding of the manual goal setting process. Then design students created some design prototypes by aligning user needs and the identified issues. The design students conducted multiple co-design sessions with end users to refine and identify potential solutions. After this, the design students created high fidelity wire-frames by incorporating end user feedback received during these sessions. In the next 12 weeks, project stakeholders were limited to development students and end users as the designers had completed their portion of the work. The developers started development of the application based on the finalised design prototype. Developers conducted usability testing sessions with users. Usability testing sessions were conducted with the participation of two developers and an end user. There were multiple usability sessions between developers and end users. During these usability sessions end users were instructed to follow a list of tasks to evaluate the usability of newly developed features of the application. One developer hosted the session by providing instructions to end users \textit{(hosting developer)} and the other developer took meeting minutes \textit{(minute taking developer)}. End users provided feedback regarding their experience during these sessions. The co-design sessions, usability testing sessions and all the other sessions were conducted online via the Zoom.  


\textit{The Case Selection:} Empathy involves nuanced human interactions. A mixed-methods approach facilitated a comprehensive understanding of these interactions. Qualitative methods, such as interviews and observations, helped us to capture the richness of individual experiences and perceptions of developers and users, while quantitative methods such as empathy test scores helped to analyse empathy trends. By combining qualitative and quantitative data, we achieved a more holistic view into the phenomenon and subsequently a more rounded form of analysis. This approach enabled triangulation, where findings from different methods were compared and validated, which enhanced the overall reliability and validity of the study. In addition, empathy involves interpersonal dynamics that may not be fully captured through purely quantitative means. An up-close and in-depth investigation, facilitated by qualitative methods, allowed us to delve into the nuances of human interactions and gain a more in-depth understanding of the empathetic processes at play.



\subsection{De-identification Process} \label{SEC:De-identification Process}
We followed a rigorous de-identification process to protect the confidentiality of participant data and to ensure participant data is untraceable. It was essential to implement a rigorous process for de-identification as there was a limited number of participants in our case study including only one female developer. In order to respect the confidentiality of our participants, we refer to them by numbers P1-P6 for developers and PU1-PU4 for users in this paper (Table \ref{TAB:Demographics of the Interview Participants}). 
We intentionally omit specific participant numbers for some quotations in the findings section to ensure participant anonymity and maintain confidentiality.

\begin{figure}[t]
    \centering
    \includegraphics[scale=.4]{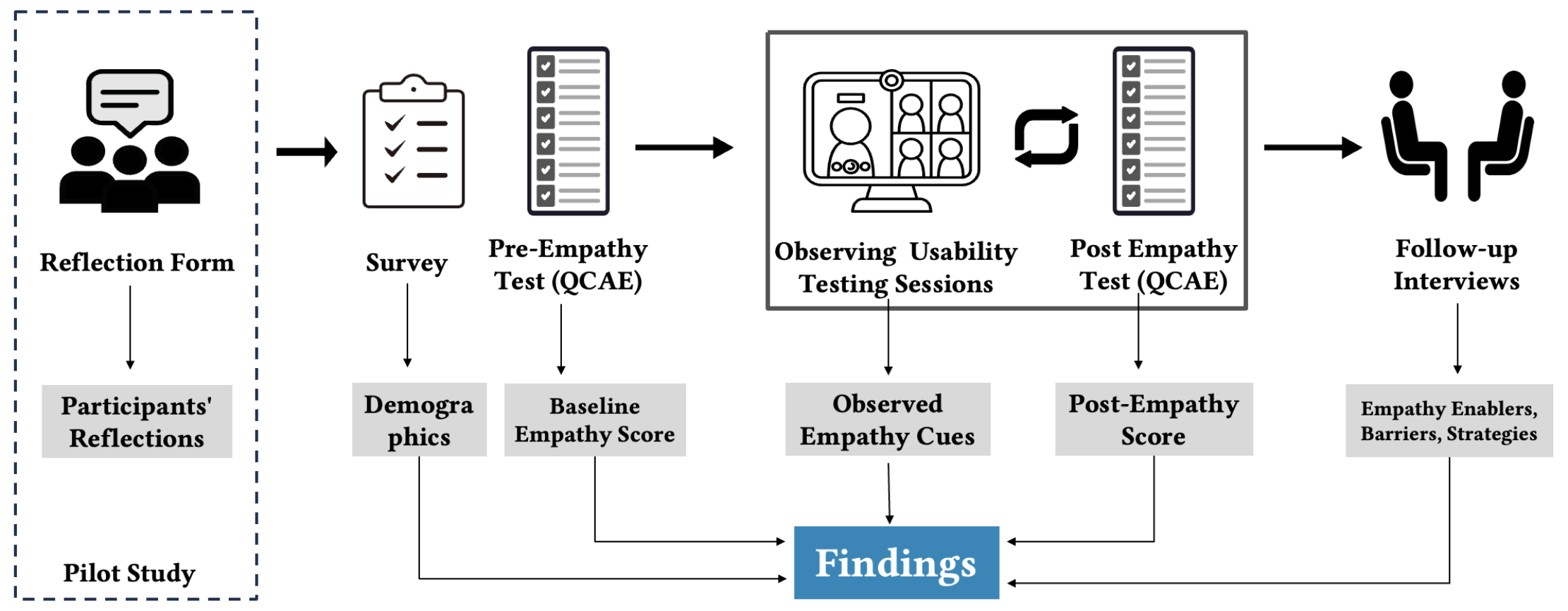}
    \caption{Overview of the research methodology}
    \label{FIG:Study Methodology}
\end{figure}

\subsection{Data Collection} \label{SEC:Data Collection}

Figure \ref{FIG:Study Methodology} shows an overview of our study methodology and outcomes of each procedure. In the first 12 weeks, we conducted a pilot study. We observed all the co-design sessions and distributed a reflection form to the participants of each session. This reflection form served as a retrospective of co-design sessions. We asked participants what went well, what didn't went well, what could be improved, and what were the surprising events happened during the sessions. 
In the next 12 weeks, we used three data collection methods: empathy test, observations of the usability testing sessions between developers and end users, and semi-structured interviews. The latter provided the most in-depth insights, helping confirm, clarify, and explain many of the insights captured from the other two methods.



   
        

        


\subsubsection{Questionnaire and Empathy Test} \label{SEC:Empathy Test}
We created two questionnaires comprising of three main types of questions: basic demographic questions, questions related to experience of working in teams and usage of eHealth apps, and an empathy test. The questionnaires were distributed among developers and end users via the Qualtrics online survey platform. Participants were instructed to complete the first questionnaire prior to their first usability session (round zero). We considered the empathy score calculated in round zero as the baseline empathy of the participants. Then the participants were instructed to complete a short version of the same questionnaire after each usability session. This short questionnaire was created by reducing the demographic and team related questions used in the original questionnaire. Along with the empathy test, we kept the questions related to the demographics and employment which had a possibility of changing during the course of the research. 

Empathy scale selection was carried out with the help of empathy experts. We considered four prominent empathy scales: Interpersonal Reactivity Index (IRI) \cite{davis1983IRI}, Empathy Quotient (EQ) \cite{baron2004EQ}, The Jefferson Scale of Empathy (JSE) \cite{hojat2001JSE} and a Questionnaire of Cognitive and Affective Empathy (QCAE) \cite{reniers2011QCAE}. We removed JSE as it has been developed specifically for health professionals and patients. We then discussed with two empathy experts (PhD in Psychology, PhD Candidate in Neuroscience Dept. of Monash Central Clinical School with special focus on empathy) to select the most appropriate scale for our study. Based on the feedback of our empathy experts -- we removed EQ as it is mainly used for assessing Autism. We then removed IRI as it seems irrelevant to our study because it assesses processes broader than empathy \cite{baron2004EQ, gunatilake2023empathy}. Despite some of the irrelevant items in its peripheral responsivity subscale, QCAE appeared to be the most appropriate empathy scale to our study. Even though there are no empathy scales designed specifically for SE \cite{gunatilake2023empathy}, we identified QCAE would be suitable to serve the purpose of our study especially due to its focus on cognitive empathy and perspective taking. 


QCAE consists of two main components and is divided into five different subscales. These two components are cognitive empathy and affective empathy. The five subscales are perspective taking which consists of ten items, online simulation which consists of nine items, emotion contagion which consists of four items, proximal responsivity which consists of four items, and peripheral responsivity which consists of four items. Altogether QCAE comprised of 31 items and participants are required to indicate how each item best describes them via a Likert scale from ``Strongly Agree'' to ``Strongly Disagree''. The scores of subscale items are summed to produce the total subscale scores. Cognitive empathy score is calculated by summing the scores of two cognitive subscales and affective empathy score is produced by summing the scores of three affective subscales. The focus of each of these subscales are explained in Table \ref{TAB:Glossary of Terms}. 

\subsubsection{Observations} \label{SEC:Observation Study}
We observed the usability testing sessions in the second 12 weeks. The purpose of our observations was to identify the empathy cues demonstrated during the interactions between software developers and users. We found a set of verbal and nonverbal empathy cues from existing literature \cite{elliott2011empathy, nicolai2007rating, chartrand1999chameleon}. Two of the authors independently observed the usability testing sessions and manually recorded these verbal and nonverbal empathy cues demonstrated by developers and users. We observed seven usability testing sessions with each session lasting approximately 45 minutes. We used a Google form to collect responses and the two observers filled out this form separately during each session. At the end of our case study, the first author compared and collated the observation notes of both observers. The first author also re-watched all the video-recorded observation sessions multiple times to record the occurrences of empathy cues (see Tables \ref{TAB:Occurrences of Verbal Empathy Cues of Developers}, \ref{TAB:Occurrences of nonverbal Empathy Cues of Developers}, \ref{TAB:Occurrences of Verbal Empathy Cues of Users}, \ref{TAB:Occurrences of nonverbal Empathy Cues of Users}). Apart from the empathy cues, we also recorded some supplementary details such as the mood of the session, how developers responded to the communication challenges of users and how developers responded to negative and positive emotions of users or vice-versa.

We implemented several measures to prevent our observations from impacting the usability sessions, aiming to maintain an unobtrusive and non-disruptive process \cite{taylor1984introduction, taylor2015introduction}. To minimise any potential influence, observers maintained a passive role by refraining from speaking and keeping their videos off throughout the sessions. Additionally, participants were informed in advance that the observations were solely for research purposes and would not use them for any other intent.


\textbf{Verbal Empathy Cues:} We observed following four verbal empathy behaviours \cite{elliott2011empathy} and these were identified as the verbal empathy cues during this study (see Table \ref{TAB:Verbal Empathy Cues}). These four verbal empathy behaviours were demonstrated in the usability testing sessions as: answers with reference to the content of the other person; content-consistent repetition of the statement of the counterpart; questioning to better understand what the other person is saying with reference to the content of the other person's statement; confirmation of the other person or showing understanding for their statement.

\begin{table*}[ht]
\begin{threeparttable}
    \footnotesize
    \centering
    \caption{Verbal Empathy Cues}
    \label{TAB:Verbal Empathy Cues}
    \begin{tabular}{ll}
        \toprule
        \textbf{Verbal Cue} & \textbf{Description}\\
        \midrule
        
        Empathic & \textit{A person responds with reference to the content of the other person.} The first behaviour is an empathic statement that\\
        Understanding & demonstrates a person has understood what the other person is saying. It is a response related to the content of the other\\ 
        Responses & person's statement. The person answers the other person, but in doing so the person refers to the content of the previous\\
        & statement of the other person. For example, the person could demonstrate this by \textit{repeating or summarizing the content of}\\
        & \textit{other person's statement when answering.}\\ 
        & \\
    
        Empathic & \textit{One person repeats the statement of the other person with the same content.} The second behaviour is verbal confirmation\\
        Affirmations & from the other person, in which the person tries to \textit{express understanding} for the other person. This behaviour is\\
        & demonstrated \textit{by summarising the content of the statement made by the other person.} In this behaviour, one person repeats\\
        &  the statement of the other person without any evaluation or answer. In this way, the first person shows that they have\\
        &  understood what the other person said.\\
        & \\
       
        Empathic & \textit{A person asks in order to better understand what the other person is saying, with reference to the content of the other person's}\\
        Conjectures & \textit{statement.} The third empathetic behaviour is a substantive question about the statement of the other person. It is a \textit{guess or}\\
        &  \textit{a question about the content of the other person's statement}, in which one person tries to understand part of the content of\\
        & the other person. When demonstrating this behaviour, the first person asks the other person a question, but this shows\\
        & that they have understood the content of what the other person said.\\
        & \\

        Empathic & A person confirms the other person or shows understanding for their statement. The fourth behaviour is expressing the\\
        Evocations & experience of the other person in different words, where the person tries to \textit{summarize the experience of the other person},\\
        & this is often done in the \textit{form of a question}. This empathy behaviour consists of affirming or showing understanding for\\
        & the other person. When demonstrating this behaviour the person refers back to the statement of the other person and the\\ 
        & person could, for example, praise the other person or could show understanding for a difficult situation of the other person.\\ 
    
        \bottomrule
    \end{tabular}
    
    \end{threeparttable}
\end{table*}

\textbf{Nonverbal Empathy Cues:} We identified facial, vocal and posture related signals as nonverbal cues. Nonverbal behaviours were identified from two existing studies \cite{nicolai2007rating, chartrand1999chameleon}. We observed the following three manifestations of these nonverbal empathy behaviours and these three manifestations were identified as the nonverbal cues during this study (see Table \ref{TAB:Nonverbal Empathy Cues}).

\begin{table}[ht]
\begin{threeparttable}
    \footnotesize
    \centering
    \caption{Nonverbal Empathy Cues}
    \label{TAB:Nonverbal Empathy Cues}
    
    \begin{tabular}{ll}
    \toprule
    \textbf{Nonverbal Cue} & \textbf{Description}\\
    \midrule
    
    Lute for careful & Involves sounds for attentive listening. This behaviour provides a verbal acknowledgement without interrupting the\\
    listening & flow of the other person's speech or speaking directly themself. For example this \textit{involves sounds like ``Mmm'', ``Ah'', ``Oh'' or}\\
    & \textit{short, monosyllabic words such as ``Ok'', ``Yes'', ``Oh yeah''.} The sounds show that an empathic person is listening to the other\\
    & person or can understand what the other person is saying \cite{nicolai2007rating}.\\
    & \\

    Nonverbal signals & Refers to the facial expressions and attitudes of the empathic person. We observed five signals for attentive listening which\\
    for attentive & consists of four signals related to facial expressions and one signal related to the attitude of the empathic person \cite{nicolai2007rating}. We\\
    listening & recorded five signals for attentive listening as \textit{multiple nods, wide open eyes \& raised eyebrows which sometimes create light}\\
    & \textit{transverse wrinkles on the forehead, a head tilted slightly to one side, the thumb \& forefinger of one hand on the chin, and}\\
    & \textit{an empathic posture where a person bends forward towards the other person and builds up slight tension in the upper body}.\\
    & This empathic posture is an indicator that the person feels addressed or actively participates in the conversation \cite{bartel2000collective}.\\
    & \\

    Similar facial & This indicates having similar facial expressions, posture, gestures of the empathic person as their counterpart. Ekman et al.,\\
    expressions, & assigned specific facial signals to four basic emotions \cite{ekman1971constants}. These emotions namely joy, anger, fear, mourning were used to\\
    posture, gestures & identify similar facial expressions. In addition to similar facial expressions, we also observed similar body postures \\
    & and similar gestures between developers and users \cite{chartrand1999chameleon}.\\
    
    \bottomrule
    \end{tabular}
    \end{threeparttable}
\end{table}

\subsubsection{Semi-Structured Interviews} \label{SEC:Interviews}
We collected data by conducting online, semi-structured interviews with ten developers and end users using open-ended questions. The interviews were approximately 30-45 minutes long. These interviews were conducted at the end of the 24-week period after all the usability testing sessions were completed. Due to the pandemic and geographical distribution of the participants, all of the interviews were conducted online via Zoom, and all were audio recorded. 

Interviews focused on the experience of the participants during the usability testing sessions, in particular what made them empathetic or what inhibited their empathy towards the other group. 
We created two interview guides for developers and end users comprised of core questions related to enablers and barriers of empathy as well as some associated questions derived from the observations. These interview guides were employed during the interviews together with appropriate follow-up questions based on the flow of the conversation. 

The questions included in the interview guides (see Appendix \ref{SEC:Interview GuidesAppendix}) are generally straightforward, with the exception of those related to the concept of empathy. Recognising the potential for ambiguity in the interpretation of empathy, we provided a clear definition of empathy (refer Table \ref{TAB:Glossary of Terms}) at the beginning of the interview to guide participants. Additionally, we maintained an open line of communication during the interviews, encouraging participants to seek clarification on any unclear or ambiguous questions. 
To further ensure the questions' clarity and appropriateness, the first author sought input from other, more experienced, co-authors. Their feedback helped in refining the wording of questions and addressing any potential confusion. Due to the nature of this case study, conducting a traditional pilot interview study was challenging, as the questions were closely tied to the project's details and participants' involvement. While a traditional pilot study was not feasible, these measures were taken to ensure the clarity and appropriateness of the interview questions within the unique context of this case study.

Further during the interviews, we didn't directly inquire about participants' empathy enablers and barriers. Instead, our approach involved prompting them with queries about instances when they perceived successful or challenging empathetic experiences. This method was chosen to allow participants to naturally share insights into their empathy without explicitly focusing on enablers and barriers. While this approach may have limitations, we believe it contributed to the authenticity of the data we gathered.

\subsubsection{Triangulation}
Three types of triangulation were used in this study \cite{runeson2009guidelines}: (a) \textit{Data source triangulation} was used in two ways: by collecting data from multiple sources such as interviews, questionnaires, observations and by collecting same data at different occasions for example administering empathy test to developers and users at different stages; (b) \textit{Observer triangulation}, by involving multiple researchers for the analysis of same data sources such as having two researchers to observe usability testing sessions and involving two researchers for the analysis of interview transcripts; and (c) \textit{Methodological triangulation}, by combining different types of data collection methods such as qualitative (interviews, observations) and quantitative methods (empathy test).

\subsection{Mixed Methods Data Analysis} \label{SEC:Data Analysis}


We used mixed methods to analyse data. \textit{Qualitative data analysis} was mainly used to analyse interview data using STGT for data analysis \cite{hoda2021STGT}. \textit{Quantitative data analysis} was mainly employed to calculate the scores of empathy test and also to analyse trends using these calculated scores. 

\subsubsection{Qualitative Data Analysis} \label{SEC:Qualitative data analysis}
We used qualitative data analysis for analysing interview and observational data. We transcribed the recordings of interviews, then stored and analysed them using NViVo software. We used STGT \cite{hoda2021STGT} to analyse interview transcriptions. STGT has been formalised as a method particularly suitable for technology-intensive domain studies such as SE \cite{graetsch2023data}. 

For this study, we specifically used the \textit{STGT for data analysis}. Our interview responses provided sufficient qualitative data for applying coding techniques but were not sufficient for full theory development. Therefore, a limited application of STGT for data analysis was found suitable for our study.
We selected this approach over other qualitative analysis methods, such as thematic analysis, due to three primary reasons: (i) our study's alignment with the socio-technical (ST) research framework proposed by STGT; (ii) the rigorous nature of STGT, resulting in original, relevant, and multidimensional outcomes; and (iii) its reflective practices, such as memoing, which facilitated layered insights and reflections. As stated before, our study aligns with the dimensions of the ST research framework proposed by STGT: ST phenomenon, ST domain and actors, ST researchers, and ST tools, techniques and data. We studied how empathy is practised between developers and users which is a ST \textit{phenomenon} because we are exploring the technical consequences of a social phenomenon i.e., empathy. The \textit{domain} of our study is SE which is a ST domain due to ``tight coupling between its social and technical aspects'' \cite{hoda2021STGT}. Also software developers and users are the \textit{actors} who play key roles in our ST phenomenon. In this study, these actors not only use their knowledge to create technology but also use their experience to increase the impact of these technological solutions. Our research team consists of ST \textit{researchers} who have requisite knowledge and skills in qualitative research, philosophical foundations, and technical experience. As ST researchers, our team has necessary technical skills and domain knowledge. The interviews were conducted by the first author who is an experienced software practitioner and an early career researcher, supported by a supervisory team with strong research and technical skills. This enabled us to understand the processes and experiences of participants. The study used several ST \textit{tools}, \textit{techniques} and \textit{data} including NVivo and Zoom. We used data collection techniques such as interviews, observations \& questionnaires to collect both qualitative and quantitative data. STGT is particularly designed to capture this type of an ST research context which is different to the native social context of traditional GT methods \cite{hoda2021STGT}. 

\begin{figure}
    \centering
    \includegraphics[width = \textwidth]{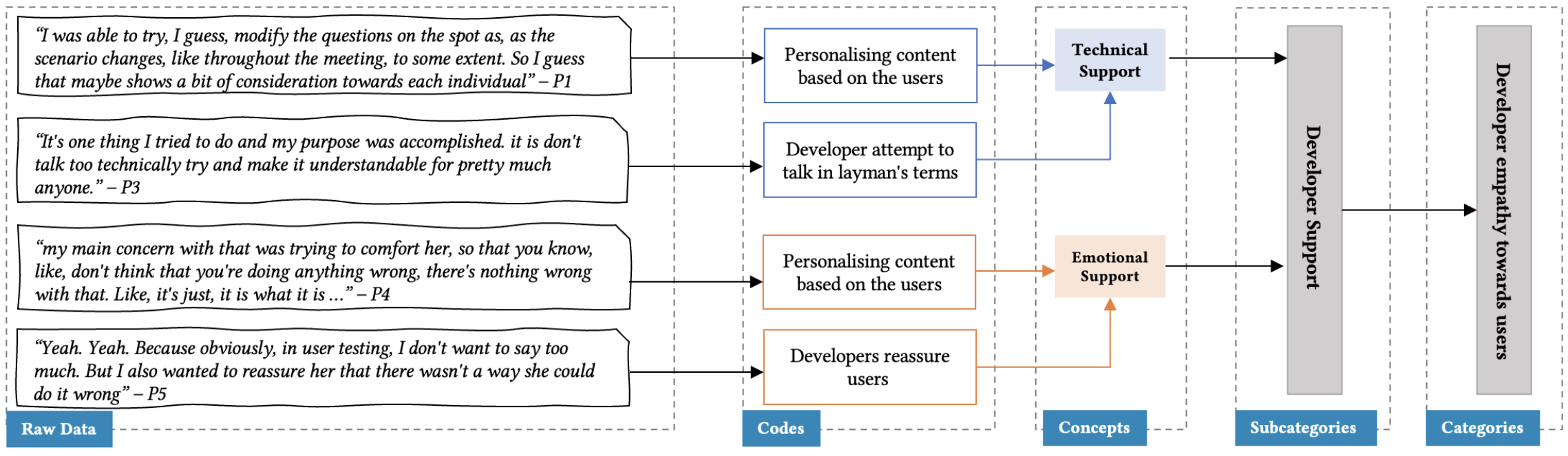}
    \caption{Emergence of the category `\textit{Developer empathy towards users}' from raw data → codes → concepts → subcategory → category through constant comparison}
    \label{FIG: STGT Example}
\end{figure}

Open coding was used to analyse the interview transcripts (raw data). The first author open coded all the interview transcripts and derived a preliminary code book. This preliminary code book was peer reviewed by third author who is an experienced grounded theorist. Based on the peer review discussion, the first author refined the preliminary code book and conducted constant comparison. Constant comparison was used to compare the codes within the same and across different transcripts. Similar codes were then grouped to form concepts, similar concepts were grouped to form subcategories and similar subcategories were grouped to form categories. An example of the STGT analysis is illustrated in Figure \ref{FIG: STGT Example}. 
For a more comprehensive example, the supplementary information package is accessible online. \footnote{https://github.com/Hashini-G/SupplementaryInfoPackage-AskPCOSStudy}

Memos are free-flowing ideas about the codes and emerging relationships \cite{hoda2011selforganizingagile}. We used basic memoing to record researcher reflections in detail. We wrote memos when we had insights about codes and their emerging relationships. Memoing also helped us to systematically document reflections on emerging concepts, subcategories, categories and possible links between them. For example, see memo in Appendix \ref{SEC:MemosAppendix}. In addition, we qualitatively matched developer and user behaviours observed during usability testing sessions to empathy cues by referring to the identified verbal and nonverbal empathy cues (Table \ref{TAB:Verbal Empathy Cues} and \ref{TAB:Nonverbal Empathy Cues}). 

\subsubsection{Quantitative Data Analysis} \label{SEC:Quantitative data analysis}
We used quantitative methods to analyse empathy test scores and represent analysed qualitative empathy cues in a numerical format. For empathy test related analysis we used measures of dispersion in descriptive statistics. Measures of dispersion describe the dispersion or the spread of data \cite{Wohlin2012Experimentation}. We used box plots to visualise data as we can easily get an overview of the data set and identify outliers. We calculated quartiles, interquartile range (IQR), maximum and minimum values of empathy test scores. This data was then used to draw box plots. We drew box plots to visualise the dispersion of baseline empathy scores, round one empathy scores and round two empathy scores of both developers and end users. These scores were visualised considering the empathy scores of five subscales of QCAE. We also illustrated total empathy scores, total cognitive scores and total affective scores of each round using box plots. 
In addition to these group based visualisations, we used line graphs to illustrate empathy scores of each individual participant. These line graphs were used to visualise individual scores in each round considering five subscales of QCAE, total cognitive, total affective and total empathy scores. 
We further analysed qualitative empathy cues by converting them to a quantitative measure by calculating occurrences of each empathy cue. We recorded how many times each of these empathy cues were demonstrated by developers and users during each usability session. This quantitative representation is explained in Section \ref{SEC:Observation Findings}.

\section{Findings} \label{SEC:Findings}


Figure \ref{FIG:Interview Findings} outlines the key findings of our interview study analysis. Through the interview analysis, we found evidence for developer awareness (Section \ref{SEC:Developer Awareness}), enablers of developer \& user empathy (Section \ref{SEC:Enablers of Empathy}) as well as barriers to developer \& user empathy (Section \ref{SEC:Barriers to Empathy}). We also found a set of potential strategies developers may use to improve empathy (Section \ref{SEC:Strategies to Improve Empathy}). We report the empathy cues identified during our observation study (Section \ref{SEC:Observation Findings}) and trends discerned from our empathy test (Section \ref{SEC:Empathy Test Findings}). We include pertinent quotations derived from our interviews that provide valuable insights into the underlying concepts. We provide references to all the participants whose interviews generated one or more codes for a particular concept. However, due to confidentiality issues we cannot include all the underlying quotations from our interviews. 

\begin{figure}
    \centering
    
    \tikzset{every picture/.style={line width=0.65pt}} 
    \footnotesize
    \begin{tikzpicture}[x=0.75pt,y=0.75pt,yscale=-1,xscale=1]
    
    \draw  [draw opacity=0][fill={rgb, 255:red, 243; green, 243; blue, 243 }  ,fill opacity=1 ] (0,214.5) -- (210,214.5) -- (210,335) -- (0,335) -- cycle ;
    \draw  [draw opacity=0][fill={rgb, 255:red, 209; green, 209; blue, 209 }  ,fill opacity=1 ] (0,194) -- (210,194) -- (210,214.5) -- (0,214.5) -- cycle ;
    \draw   [color={rgb, 255:red, 74; green, 74; blue, 74 }  ,draw opacity=1 ] (0,194) -- (210,194) -- (210,335) -- (0,335) -- cycle;
    
    \draw  [draw opacity=0][fill={rgb, 255:red, 243; green, 243; blue, 243 }  ,fill opacity=1 ] (0,355.5) -- (210,355.5) -- (210,450) -- (0,450) -- cycle ;
    \draw  [draw opacity=0][fill={rgb, 255:red, 209; green, 209; blue, 209 }  ,fill opacity=1 ] (0,335) -- (210,335) -- (210,355.5) -- (0,355.5) -- cycle ;
    \draw   [color={rgb, 255:red, 74; green, 74; blue, 74 }  ,draw opacity=1 ] (0,335) -- (210,335) -- (210,450) -- (0,450) -- cycle;
    
    
    \draw  [draw opacity=0][fill={rgb, 255:red, 243; green, 243; blue, 243 }  ,fill opacity=1 ] (210,214.5) -- (462,214.5) -- (462,335) -- (210,335) -- cycle ;
    \draw  [draw opacity=0][fill={rgb, 255:red, 209; green, 209; blue, 209 }  ,fill opacity=1 ] (210,194) -- (462,194) -- (462,214.5) -- (210,214.5) -- cycle ;
    \draw   [color={rgb, 255:red, 74; green, 74; blue, 74 }  ,draw opacity=1 ] (210,194) -- (462,194) -- (462,335) -- (210,335) -- cycle;
    
    \draw  [draw opacity=0][fill={rgb, 255:red, 243; green, 243; blue, 243 }  ,fill opacity=1 ] (210,355.5) -- (462,355.5) -- (462,450) -- (210,450) -- cycle ;
    \draw  [draw opacity=0][fill={rgb, 255:red, 209; green, 209; blue, 209 }  ,fill opacity=1 ] (210,335) -- (462,335) -- (462,355.5) -- (210,355.5) -- cycle ;
    \draw   [color={rgb, 255:red, 74; green, 74; blue, 74 }  ,draw opacity=1 ] (210,335) -- (462,335) -- (462,450) -- (210,450) -- cycle;
    

    \draw  [draw opacity=0][fill={rgb, 255:red, 5; green, 136; blue, 188 }  ,fill opacity=1 ] (143,315) -- (277,315) -- (277,360) -- (143,360) -- cycle ;
    \draw  [draw opacity=0][fill={rgb, 255:red, 74; green, 74; blue, 74 }  ,fill opacity=1 ] (0,194) -- (34,194) -- (34,214.5) -- (0,214.5) -- cycle ;
    \draw  [draw opacity=0][fill={rgb, 255:red, 74; green, 74; blue, 74 }  ,fill opacity=1 ] (428,194) -- (462,194) -- (462,214.5) -- (428,214.5) -- cycle ;
    \draw  [draw opacity=0][fill={rgb, 255:red, 74; green, 74; blue, 74 }  ,fill opacity=1 ] (0,335) -- (34,335) -- (34,355.5) -- (0,355.5) -- cycle ;
    
    \draw  [draw opacity=0][fill={rgb, 255:red, 74; green, 74; blue, 74 }  ,fill opacity=1 ] (428,335) -- (462,335) -- (462,355.5) -- (428,355.5) -- cycle ;

    \draw (155,317) node [color={rgb, 255:red, 243; green, 243; blue, 243 }  ,opacity=1 ][anchor=north west][inner sep=0.75pt]   [align=left] {\begin{minipage}[lt]{80pt}\setlength\topsep{0pt}
    \begin{center}
    \normalsize \textbf{Empathy in Developer - User Interactions}
    \end{center}
    \end{minipage}};
    
    
    \draw (2,219) node [anchor=north west][inner sep=0.75pt]   [align=left] {\textbf{Developer awareness of users}\\\textit{Emotional: User needs, Concerns, Reactions, Feelings} \\\textit{Technical: User understanding, Technical proficiency,} \\\textit{Importance of empathy}};
    \draw (65,204.75) node   [align=left] {
    \begin{minipage}[lt]{100pt}\setlength\topsep{0pt}
    \begin{center}
    \textbf{\small Awareness}
    \end{center}
    \end{minipage}};
    \draw (2,270) node [anchor=north west][inner sep=0.75pt]   [align=left] {\textbf{Developer self-awareness}\\\textit{Technical jargon, Differences in empathy, Less} \\\textit{developer awareness of user's statements}};
   
    \draw (2,365) node [anchor=north west][inner sep=0.75pt]   [align=left] {\textbf{Developer support}\\\textit{Technical support, Emotional support}};
    \draw (65,345) node   [align=left] {\begin{minipage}[lt]{100pt}\setlength\topsep{0pt}
    \begin{center}
    \textbf{\small Strategies}
    \end{center}
    \end{minipage}};
    \draw (2,401) node [anchor=north west][inner sep=0.75pt]   [align=left] {\textbf{Prospective developer acts of empathy}\\\textit{Better preparation, More follow-up questions, Slow}\\\textit{down the session, Support users more, Taking time}\\\textit{to implement user suggestions}};

    
    
    
    \draw (212,219) node [anchor=north west][inner sep=0.75pt]   [align=left] {\textbf{Enablers of developer empathy}\\\textit{Reflection in action, Commonality, Familiarity, Understanding,}\\\textit{Appreciation, Connection, Interaction, Awareness, User confusion,}\\\textit{Involvement, Gender similarities, Similar level of technical ability}};
    \draw (212,270) node [anchor=north west][inner sep=0.75pt]   [align=left] {\textbf{Enablers of user empathy}\\\textit{Familiarity \& Connection, Connection to app's goal, Appreciation,} \\\textit{Understanding of developer thought process, Understanding of} \\\textit{developer needs, Mode of communication}};
    \draw (400,204.25) node   [align=left] {\begin{minipage}[lt]{100pt}\setlength\topsep{0pt}
    \begin{center}
    \textbf{\small Enablers}
    \end{center}
    
    \end{minipage}};
    \draw (212,365) node [anchor=north west][inner sep=0.75pt]   [align=left] {\textbf{Barriers to developer empathy}\\\textit{Less interest, Unfamiliarity,  Limited time for reflection in action,} \\\textit{Less interaction, Difficulty to resonate, More concentration on the} \\\textit{task, Poor connection with users, Developer difficulties}};
    \draw (212,416) node [anchor=north west][inner sep=0.75pt]   [align=left] {\textbf{Barriers to user empathy}\\\textit{Poor connection, User nervousness, Users feeling awkward,} \\\textit{Confusion on terminology}};
    \draw (400,345) node   [align=left] {\begin{minipage}[lt]{100pt}\setlength\topsep{0pt}
    \begin{center}
    \textbf{\small Barriers}
    \end{center}
    
    \end{minipage}};
    \draw (2,199) node [anchor=north west][inner sep=0.75pt]  [color={rgb, 255:red, 232; green, 232; blue, 232 }  ,opacity=1 ] [align=right] {\textcolor[rgb]{0.93,0.92,0.92}{{\large 5.3}}};
    \draw (442,199) node [anchor=north west][inner sep=0.75pt]  [color={rgb, 255:red, 232; green, 232; blue, 232 }  ,opacity=1 ] [align=right] {\textcolor[rgb]{0.93,0.92,0.92}{{\large 5.4}}};
    \draw (2,339) node [anchor=north west][inner sep=0.75pt]  [color={rgb, 255:red, 232; green, 232; blue, 232 }  ,opacity=1 ] [align=left] {\textcolor[rgb]{0.93,0.92,0.92}{{\large 5.6}}};
    \draw (442,339) node [anchor=north west][inner sep=0.75pt]  [color={rgb, 255:red, 232; green, 232; blue, 232 }  ,opacity=1 ] [align=left] {\textcolor[rgb]{0.93,0.92,0.92}{{\large 5.5}}};
    
    \end{tikzpicture}

    \caption{Overview of Key Interview Findings derived using STGT Analysis.}  
    \label{FIG:Interview Findings}
\end{figure}

\subsection{Developer Awareness} \label{SEC:Developer Awareness}
It became apparent through our data analysis, that awareness in terms of \textit{developer awareness of users} and \textit{developer self-awareness of own actions and emotions} was a first step towards enabling empathy.


\subsubsection{Developer Awareness of Users}
From the answers of developers, we noticed that they had a good \textit{emotional and technical awareness of users}. When considering emotional awareness, developers noticed both \textit{positive and negative user emotions}, and were more concerned about negative user emotions. 
We asked developers about the moments that they enjoyed the most during the sessions. Developers stated that they enjoyed the session introduction more as they felt users were more relaxed during that period compared to the rest of the session [P1, P3].

Occasionally, developers sensed that users might have felt like they were being tested during the usability sessions \textit{[negative emotions]}. Developers said that it was not nice to observe such user emotions as the underlying purpose of conducting these sessions was only to measure the usability of the app and not the capabilities of users [P5]. When asked, developers noted that they could have empathised better with users when the users were seen to be stressed or under pressure \textit{[negative emotions]}. However, developers faced a dilemma of whether to act on their empathy by helping users resolve their confusions or to let them struggle and find their own way, which would allow elicitation of unadulterated user feedback on the software. Developers stated that users had quite amusing reactions \textit{[positive emotions]} when they explained the purpose of the session, which is to only test the app and not the users. Developers said that users seemed much comfortable when the developers explained there were no right or wrong answers. Overall, they were able to resonate with user emotions and reasons for these emotions based on how the users acted, despite not being able to fully understand all their concerns regarding the app [P3, P6].

\begin{quote}
\small
    \faIcon{comments} \textit{``...in user testing, I don't want to say too much. But I also wanted to reassure [user] that there wasn't a way [user] could do it wrong. Like we weren't testing [user]. But I think [user] felt like that, which wasn't that nice.'' - P5}
    
\end{quote}


As observers of these usability sessions, we noticed developer remarks to users on not testing user spellings \textit{(``it's okay, we are not testing your spellings'')} when users apologised for entering the wrong spellings. When we asked developers about the intention of their remarks, developers explained their genuine intention was to inform users that they don’t have to be perfect. However, reflecting on these remarks developers stated that their words might have sounded judgemental from a user’s perspective [P2]. This suggests that developers were considerate about user feelings, not just during the usability sessions, but even when reflecting on their remarks after these sessions. 



Developers had a good understanding of \textit{users' needs and concerns}. Developers stated that they understood the \textit{need} of the users to have options to check-in for goals both daily and weekly. Developers stated this was due to user's ADHD (Attention-deficit/hyperactivity disorder) condition that made user well focused on some days and out of focus on other days. Even without having ADHD, they could imagine how users would feel in those instances [P3]. This may be an instance where the developers exhibited \textit{perspective taking}. Another developer said that the interaction during usability sessions helped them to understand user confusions \textit{[concerns]}. The developer said that they understood the reasons for such user confusions by observing mouse movements of users, listening to users' explanations and where they are leaning towards in those situations, enabling their empathy towards users [P4]. 


It was clear that the developers were \textit{aware of users’ technical competence}. Some users were seen to be technically proficient [P2] while others suggested perceived gaps in technical understanding that did not hinder the tasks and app use [P6]. P4 said having family members who were not very tech savvy allowed them to understand their users, demonstrating \textit{perspective taking}.



Users listed difficulties in understanding some terms used in the app. However, they didn’t want to discourage the developers and make them feel bad about their app [PU3]. This indicates some evidence of users' awareness of developer emotions which can be a gesture of goodwill and/or an indication of the connection they share.

\subsubsection{Developer Self-Awareness of Own Actions \& Emotions}
When asked, one developer stated that they thought there was not much to empathise with end users in this particular project where the only connection between developers and users was during the usability sessions [P5]. However, upon continuing the discussion on the potential acts of empathy, the same developer stated that they were able to empathise with end users fairly well. Another developer stated that they didn’t observe any significant user issues during the usability sessions, hence they didn't feel the need to empathise with users [P6]. Having said that, the same developer stated that they were able to understand users' perspective and feelings. The developer said that being in a team also helps with empathising as a whole, recognizing that different individuals empathise in distinct ways. On the contrary, all the other developers stated that they were able to empathise with users [P1-P4] without doubting the need for expressing empathy. This suggests that some developers might not have a proper understanding of empathy or might not be aware of the need of empathy towards users or the impact of empathising. 

\begin{quote}
\small
    \faIcon{comments} \textit{``...I don't know if there was much to empathise with the customers about in this particular project. Because the only contact I really had with them was user testing. It was more just observing.'' - P5}
\end{quote}


Developers demonstrated their awareness of using technical jargon when inquired about users’ understanding of explained technical aspects. Developers acknowledged that they might have occasionally delved into too much technical detail, but they rephrased their explanations upon noticing user confusion. Upon reflecting on the use of technical jargon, developers said that they hadn't given it much thought at the time. However, developers stated that they could have provided better explanations to enhance user understanding [P1, P2]. When asked developers about being unresponsive to some user apologies, they explained that it happened because they were distracted due to some unexpected events [P2]. 
This suggests that sometimes developers were well aware of their actions and behaviours, and tried their best to accommodate the users. However, developers occasionally lacked self-awareness, i.e., how some of their actions impact the users and, developers realised it only when inquired during the interviews.

\subsection{Enablers of Empathy} \label{SEC:Enablers of Empathy}
Both developers and users identified several enablers of their empathy. Enablers of developer empathy include: \textit{Reflection in action, Commonality, Familiarity, Improved user understanding, Awareness of user actions and emotions, Appreciation, User confusion, Developer involvement, Technology literacy related issues, Gender similarities, Better connection, User interaction and Similar technical ability}. Enablers of user empathy include: \textit{Mode of communication, Understanding of developer thought process, Understanding of developer needs, Developer familiarity and connection, Connection to the app's goal and appreciation of developer efforts}. 
We describe these enablers in detail in the following subsections, emphasising how they led to developer and user empathy.

\subsubsection{Enablers of Developer Empathy}
There were different enablers of developer empathy towards users. Developers explained their perceptions of these different enablers of empathy when asked about how well they were able to empathise with users. We asked about their views on any special reasons for their empathy and how meeting roles impacted their ability to empathise. We also posed questions derived from our observations, such as why they felt more connected to some users over others. We mapped the demonstrated behaviours and expressed ideas to different empathy types such as cognitive and emotional empathy. We also mapped the empathy cues to the ideas expressed in interviews.

Developers stated that \textit{commonality} is one of the enablers of empathy [P2, P4]. App feature reasoning was indicated by a developer as a key enabler of their empathy. Developers stated that talking about the app itself and providing suggestions regarding potential new features was one of the instances where they felt a commonality with the users [P2]. Another developer described that meeting users with a similar personality and same background (both were university students) were the sources of commonality which reinforced empathy [P4]. The developer further stated that the technical issues users faced were also another form of commonality as the developer had experience in solving these technical issues with their family members who are not very familiar with technology.
\begin{quote}
\small
    \faIcon{comments} \textit{``I think I did empathise with them. I did understand some of the problems they have. And I understood where they came from with the point of view that they came from. Because from my understanding is that like, I guess my observation was that most of them that I did the interviews with, they were mostly not technology like knowledgeable I guess, as I am. So I could see it most of the time, because I have family members who are also not very technologically savvy and stuff like that. So yeah, I think I did a pretty good job of understanding what they were like feeling and like using, how to feel it and how they acted when they were using the app.'' - P4}
\end{quote}

\textit{Gender similarities} were also an enabler of empathy. A developer indicated that their gender might have assisted in understanding the thought process of the users and empathising with them. 
\begin{quote}
\small
    \faIcon{comments} \textit{``...I think that they were women probably contributed to me being able to empathise with them more, because I could maybe relate to what their thought process bit better'' - Anonymous for participant privacy}
\end{quote}

\textit{Observing user confusion} was another enabler of empathy [P2, P4, P6]. A developer stated that users had difficulties in interpreting some questions. Users were occasionally confused by the terminology used in the app and the instructions provided during the usability sessions [P6]. The developer added that it was natural for first time users to feel confused about both app and usability testing process, as everything was completely new to them. Another developer admitted that the app would have been confusing to the users as they do not have access to all the information developers had while developing the app [P2] which reinforced the statement of P6. The developer further stated that it was easier for them to see these confusions from the users’ perspective. Developers stated that the direct interaction with users helped them to understand the confusing features and the users’ behaviour in such instances [P4]. Developers also stated that this helped them to understand users’ perspective. Both P2 and P4 stated that user confusions assisted in better understanding the users’ perspective which exhibits the practice of \textit{perspective taking} which comes under the umbrella of \textit{cognitive empathy}. 

\begin{quote}
\small
    \faIcon{comments} \textit{``...I was more easily able to see from like her perspective, it would be confusing, because she doesn't have all the information that I have, as I was making it.'' - P2}
\end{quote}

\textit{Technical issues} and having a \textit{similar technical proficiency} were also triggers of developer empathy [P6]. Developers expressed their ability to empathise with users' technical difficulties, drawing from their personal encounters with technical issues while using other apps. Developers stated that they were more empathetic towards technical issues encountered by technically competent users, as they believed they could personally face similar confusion if they were using the application. We could argue that this is another instance where developers demonstrated \textit{cognitive empathy} by trying to put themselves in the shoes of the users. 


Developers stated that different meeting roles helped them with \textit{reflection in action}, another enabler of empathy [P2, P5]. A developer stated that it was easier for them to reflect and understand user feedback while taking the meeting minutes [P2]. Another developer stated that it was easier to reflect while hosting the session, as there was an opportunity to ask follow-up questions to clarify user responses [P5]. 

\begin{quote}
\small
    \faIcon{comments} \textit{``... I think maybe in the session with [user], I empathised with [user] more. Maybe mostly because I was the one running it. So I was able to ask the questions that were coming into my head. Whereas the one where [Other developer] was hosting it, [Other developer] wasn't following up with questions that I was thinking of. So that was kind of a bit less of a connection for the two of us.And I think it was easy to understand what they were trying to say when I was hosting because I could pull up with different questions.'' - P5}
\end{quote}

Developers expressed their ability to understand user thoughts [P2], confusions [P4], and feelings when using certain features due to different conditions like ADHD [P3]. Developers believed that this \textit{improved user understanding} helped them to empathise with users [P2-P4]. This may also be seen as an act of \textit{cognitive empathy}. Developers found it easy to engage with the users. Developers emphasised that users excelled in both task performance and providing feedback. Developers stated that their \textit{awareness of user actions and emotions} helped them to be more empathetic towards users [P3]. This may also be a way of displaying \textit{emotional empathy}. 

\begin{quote}
\small
    \faIcon{comments} \textit{``when we're doing some of the things like check ins and [user] said that [user] had ADHD. And [user] was describing like, sometimes [user] is really focused, [user] wants to check in every day. And if [user] is not really focused, [user] wants to check in only once a week. I don't have ADHD, but I could imagine how [user] would feel like in those instances, where [user] would feel the urge to be really regular with it, and then sometimes not want to be at all.'' - P3}
\end{quote}


\textit{User appreciation} was another trigger of developer empathy. Developers expressed appreciation for users, as they were nice, helpful, and sacrificed their time to support the developers [P3, P5]. Developers said that this sense of appreciation boosted their ability to understand and empathise with users. This also corroborates one of the verbal empathy cues we found in the literature which is praising the other person or showing understanding for a difficult situation of the other person (Verbal empathy cue 4).
Developers said that hosting helped them to be more involved in the usability sessions and this \textit{involvement} aided them to empathise with users [P3]. \textit{User interaction} was another enabler of empathy. Developers highlighted that the form of user interaction in these sessions allowed them to focus on users' thoughts, needs, and feedback, which in turn, boosted developer empathy [P1, P6]. 

\begin{quote}
\small
    \faIcon{comments} \textit{``being able to interact with people in that way, means that you're not focusing on yourself and trying to make sure that you're performing up to a task, like you're actually able to pay attention to what the end users want and the way that they're thinking about it...'' - P6}
\end{quote}

\textit{Familiarity} is another enabler of developer empathy. Observed developer familiarity can be divided into two groups as \textit{closeness to code} and \textit{familiarity of user issues}. Developers found it easier to empathise with user feedback when they were familiar with a particular feature. This familiarity allowed them to ask insightful follow-up questions, improving their understanding of the user [P2]. Developers stated that their prior awareness of less user-friendly features allowed them to empathise more with users who encountered difficulties with these features [P4]. This suggests that the developers’ closeness to code enabled them to view things from users’ perspective and be more empathetic towards them. Developers pointed out that they had personally experienced some of the user issues with their family members, and this familiarity increased their empathy towards users [P4]. 

\begin{quote}
\small
    \faIcon{comments} \textit{``... I was really easily able to identify with like the use of the software. Like whenever they had feedback, or thought things weren't working properly, I was really easily able to empathise with, like, yeah, they use the program, which I've been working on a lot. I suppose that's just because I was really familiar with it. Like I could see that the flaws in it, which they pointed out.'' - P2}
\end{quote}

Developers said that spending more time with users helped them build a \textit{strong connection}, which improved their empathy towards users [P4]. Developers stated several reasons for establishing a good connection with users. They found it easier to connect with highly engaged users, and their first-ever usability session also fostered a strong connection [P1]. Developers also expressed that the ease of engaging with users [P1, P6] was another contributing factor. Some developers even noted that it was easier to engage with users than they had initially anticipated. 
Users also supported this idea by providing positive feedback. They found it easy to engage with developers and had no suggestions for changes in \textit{user engagement}, even for future sessions [PU1, PU2, PU3]. This suggests that users were pleased with the experience they had while engaging with the developers. 
Developer involvement and familiarity were also significant factors in establishing a strong connection with users. Some developers felt a stronger connection when hosting sessions due their ability to build a rapport [P3,P4]. In contrast, others found a strong connection while taking minutes due to reflection in action [P1, P2]. 
In terms of familiarity, developers stated it would have been easier to understand and converse with more familiar users [P4, P5]. They also found it easier to connect with users who shared similar technological interests [P4, P6]. These similar interests served as a shared experience that strengthened their connection with users.
This indicates that there were many instances where developers established a strong connection with users, and this connection further reinforced their empathy towards users. We also observed a positive correlation between empathy and connection.

\begin{quote}
\small
    \faIcon{comments} \textit{``... Yeah, I think it's just like, the longer time you spend with them [users], and the more of them that you as the interviewer, like,  the ``developer'' was able to interact with the participant, I felt like the better it got, the more empathy I felt like I had with the like, the more closely like, I guess, you know, related, I guess it's the wrong word. Like, I felt like more like a connection with them even more than what I interacted with them.'' - P4}
\end{quote}

\subsubsection{Enablers of User Empathy}
All users stated that they empathised with developers and understood their intentions, indicating their self-awareness of empathy towards developers [PU1-PU4]. They also described enablers that made them empathetic towards developers. In addition, users explained certain positive actions that might have contributed to their empathy towards developers. 
Ability to \textit{communicate with developers face-to-face}, rather than through email [PU1] was an enabler of user empathy. Users stated that they \textit{understood developers' thought processes} by interacting with the actual app, which was another enabler of user empathy [PU1].


Users stated that \textit{understanding developers' needs} made them empathetic towards developers. Users described that they understood the information developers were seeking and the desired outcomes. This understanding allowed them to empathise with developers' requirements. Users stated that they provided the best possible feedback to developers. Users considered this as a way to support the developers in expanding the app's development and exploring new opportunities within their work [PU2, PU3]. Users were comfortable to give their blunt feedback since developers assured them not to be concerned about hurting their feelings [PU2]. \textit{Empathy due to understanding of developers' thought processes and needs,} demonstrates perspective taking, which is encompassed under the umbrella of cognitive empathy.

Users empathised with developers due to their \textit{connection to the app’s goal}. They found it exciting to be a part of a project that transformed an idea into a useful app particularly for females like them. Users believed the app would have a positive impact on their lives and those of other PCOS patients they knew. Users believed developers would also have empathy towards PCOS patients like them as the they built this app to support such patients. Users stated this realisation helped with their empathy [PU3]. 


\begin{quote}
\small
    \faIcon{comments} \textit{``I could empathise with them because I have PCOS and I could see that it's an app that would be helpful to myself and the people that I know. And so for that reason, I felt like there was a personal connection to the app. And I could understand that if they were trying to develop an app that would help people then they themselves would have empathy or an understanding that this has an impact on women. And so for that reason, I felt like I could empathise with that.'' - PU3}
\end{quote}

Users were empathetic due to \textit{appreciation of developer effort}. Users understood the developers and recognised their dedication to create an app that would be beneficial for them [PU2]. Users recognised developers’ persistent passion through their commitment to the app [PU3]. Users also understood developers sought user feedback to improve the app and understood developers' intentions [PU4]. Users were appreciative towards developers due to these reasons and it also became an enabler of their empathy. 

\begin{quote}
\small
    \faIcon{comments} \textit{``I was able to sort of empathise with them, because I know that they're working hard to make, like a program that would help someone like me. So that in itself was just like, okay, I get where you're coming from. And I want this too so I'm kind of like, yeah, just understood them for just that.'' - PU2}
\end{quote}

Users believed having usability sessions with the same developers would strengthen their empathy and connection with the developers due to \textit{familiarity and established connection} [PU3, PU4]. 

\begin{quote}
\small
    \faIcon{comments} \textit{``I feel previously connected to them and having liaison with them, I thought, or I would think if it's the same type of research, or, you know, maybe because they're the same type of people in that environment, they know what's going on. Maybe it might be the same. I think I'm just comfortable with doing it...'' - PU4}
\end{quote}

Users explained positive behaviours that may have enabled their empathy towards developers. 
Users were excited to join the usability sessions, demonstrating a clear understanding of the developers' needs and providing honest feedback [PU2]. Users found the sessions enjoyable and relaxed. The casual conversation with developers at the start helped users to be more relaxed. Users also appreciated the developers' patience and respectful engagement. They felt connected with the developers during all the sessions, allowing them to comfortably express their thoughts, as the developers were receptive to their feedback. The developers' clear communication helped users to engage with them successfully [PU3]. These user responses regarding their engagement with developers signify a positive experience for users during these interactions. This suggests that users had a positive engagement with developers, which could have fostered empathy towards the developers.

\begin{quote}
\small
    \faIcon{comments} \textit{``Yeah, I think they were extremely patient with me. I'm not very good with technology. And so for some of the different aspects of the app that they were showing me they were very patient and could communicate with me in a way that I understood. So they showed me that kind of sensitivity around that, which was good.'' - PU3} 
\end{quote}

Users provided \textit{positive feedback about usability sessions} and \textit{the app}. With regards to usability sessions, users appreciated the developers' flexibility in scheduling [PU2]. Users suggested having clearer instructions on the type of device to use in the sessions would be nice, but otherwise, they were content with the usability sessions [PU4]. With regards to the app, users enjoyed speaking out loud while using the app. They believed the app will be highly useful upon completion [PU1]. Users were also pleased with their ability to provide feedback and found the practical side of using the app enjoyable [PU2, PU3]. Users were excited to see the app for the first time and found it to be user friendly [PU3]. Users were able to reflect on their first session and provide more authentic feedback in the second session [PU3]. Users even suggested some new features they would like to have in the app [PU2].  

\begin{quote}
\small
\faIcon{comments} \textit{``I think it's gonna work. Like once the app is like, complete and stuff, I think it's gonna work really good. That's like going through like the first part of the study. And this part was good to get an insight as well into what goes on behind making that sort of thing.'' - PU1}
\end{quote}

User feedback about the app and usability sessions suggests that users enjoyed the experience they had while using the app. This indicates that there is a possibility that these users will use the app once it is released which could be an indication of the usability of the app. All these responses show the positive actions of users towards the interaction with the developers that may have reinforced their empathy.

\subsection{Barriers to Empathy} \label{SEC:Barriers to Empathy}
Both developers and users explained barriers that they perceived to their empathy with the other. Key barriers to developer empathy include: \textit{Less developer interest, Task centredness, Unfamiliarity of features, Limited time for reflection in action, Difficulty in understanding user struggles, Difficulty in empathising with low-technology literacy related user concerns, Less interaction and Difficulty to resonate with collective experiences of women.} Major barriers to user empathy include: \textit{Poor user connection with developers and their Negative emotional responses.} 

\subsubsection{Barriers to Developer Empathy}
Developers perceived \textit{less interaction with users} as a barrier to their empathy. During the minute taking (MT) role, they couldn't ask follow-up questions, which hindered their connection with users as the hosting developer did not address these questions either [P5]. Developers could not build a rapport with the users during MT and it made understanding users difficult. Developers felt like mere observers during MT and believed only hosting developers should interact with users because of the session format [P4]. According to developers, the fast pace of some sessions may have hindered their empathy [P1]. 

\begin{quote}
\small
    \faIcon{comments} \textit{``...Whereas the one where [other developer] was hosting it, [other developer] wasn't following up with questions that I was thinking of. So that was kind of a bit less of a connection for the two of us...'' - P5}
\end{quote}

\textit{Unfamiliarity of users and hosting sessions, and limited interaction due to MT} were the major reasons for poor developer-user connection [P2-P4]. Sometimes developers were unfamiliar with the information shared by users, which made it difficult to continue the conversation. Also hosting felt awkward initially as it seemed more like giving instructions than a real conversation. Developers found the MT role to be more of a background role, limiting their ability to interact with the users. These factors made them less connected to the users. 

Some developers stated that if they were less involved and less interested they would not empathise much with the users. This indicates that \textit{reduced developer interest can lead to less empathy towards users} [P1]. Developers explained their \textit{task centredness} made them less focused on users [P1, P4]. 

\begin{quote}
\small
    \faIcon{comments} \textit{``...I think I was more focused on trying to present the tasks as clearly as possible...'' - P1}
\end{quote}


Developers empathised more with user feedback when they were familiar with the software features, which demonstrates empathy due to closeness to code. However, developers were \textit{unable to empathise with user feedback much when the features were unfamiliar} to them. They elaborated that when a feature was familiar to them, they had already considered various ways to implement it. But if it was unfamiliar, they hadn't explored these options, which made them less empathetic towards user suggestions [P2].

\begin{quote}
\small
    \faIcon{comments} \textit{``.. I was able to empathise with their feedback a lot more, than on features where I wasn't as familiar because I wasn't, I hadn't already thought through, like, all the different options of how we could have done it for example, whereas on stuff that I was really familiar with, I may have already like thought of that.'' - P2}
\end{quote}

Developers stated that \textit{limited time for reflection in action} was a barrier to their empathy. They did not have enough time to think about user feedback while hosting, hence they were unable to empathise with those suggestions [P2]. 
\begin{quote}
\small
    \faIcon{comments} \textit{``..Whereas when I was hosting it, it was very like, you know, you just talk and you don't really have time to stop and like think about that a specific feedback.'' - P2}
\end{quote}

\textit{Reflection on action} was also a barrier to developer empathy. Sometimes developers could not properly understand what users meant when they reflect on user feedback [P2]. Developers thought that asking follow-up questions would have been helpful in these instances. 
Developers stated they had a \textit{difficulty in understanding some of the user suggestions} and they struggled to empathise with such suggestions. Developers stated that it was hard for them to understand what the users were envisioning and they could not get on the same page with users for all of their suggestions [P2].

\begin{quote}
\small
    \faIcon{comments} \textit{``Well, I think like there are other suggestions where I didn't empathise as well. So when user wanted to have like a tick and flick, or one of the features and, like, it was, I mean, it's hard to say like, I struggled to empathise, because I just didn't really understand.''- P2}
\end{quote}

Developers had \textit{difficulties in understanding user struggles} because they had not personally experienced these issues. This lack of personal experience hindered their empathy. Developers found it challenging to empathise with user issues that were unique to less tech-savvy users [P6]. Since developers were well-versed in technology and familiar with the app, they couldn't relate to these particular \textit{low-technology literacy related user concerns} [P5, P6]. Developers had a \textit{difficulty in finding the balance of technical expression}. They believed that excessive explanations could confuse users and make it hard to obtain relevant user feedback. However, developers felt that they found a balance and didn't delve into too much technical details [P6].


\begin{quote}
\small
    \faIcon{comments} \textit{``.. I felt that like I didn't really experience the problems they were having per se, like, we kind of like show them what we have. And they were like, yeah, it looks good, good job. And like, obviously, they had maybe some small issues using or whatever, but there wasn't anything significant. So it was hard to you know, really empathise with people who didn't really feel like we're struggling. I didn't feel the need to empathise with them...'' - P6}
\end{quote}


Developers stated that \textit{difficulty to resonate with collective experiences of women} was another barrier for their empathy. However, some developers argued that they would be able to empathise the same way even without having the same gender. 
\begin{quote}
\small
    \faIcon{comments} \textit{``I think in this project, one thing that made it a little bit more difficult is just gender. Like, where end users are exclusively women. As as a man, it's like, already, at least one step out of being able to empathise fully. It doesn't change, like how much I'm trying to, but I think it changes the reality of like, there are collective experiences that women have that I just don't know, haven't experienced personally. And from that perspective, it definitely makes it more difficult to fully empathise in the way that maybe you will be able to have a stronger connection with..'' - Anonymous for participant privacy}
\end{quote}

\subsubsection{Barriers to User Empathy}
\textit{Poor user connection with developers, and users’ negative emotional responses}, were the two major barriers to user empathy. Users had a \textit{poor connection} with the developers who played the less interactive role of MT. Even though users understood the role of these developers, they perceived a lack of connection with them [PU4]. 

\begin{quote}
\small
    \faIcon{comments} \textit{``I have nothing against [minute-taking developer] or anything. It's just, you know, I didn't have much interaction with [minute-taking developer] at all. That's fine. Oh, definitely. because that I feel like well, previously, I did the other session at first, and then I was invited back. So I, for example, you know, I was aware of [this developer] previously, but again, I still hadn't had much to do with [this developer]..'' - PU4}
\end{quote}

Users' \textit{negative emotional responses}, such as nervousness, confusion, and discomfort, were the other major empathy barriers. Users described various reasons for their nervousness during usability sessions. They stated feeling rushed due to \textit{confusion} about the meeting time and work commitments, which added to their nervousness. Additionally, their \textit{unfamiliarity} with what to expect during the sessions contributed to their nervousness. Users also expressed a \textit{strong desire to provide high quality feedback} to developers, which increased their nervousness as they didn't want to disappoint developers. Lastly, the \textit{anticipation of the questions} that might be asked during the sessions heightened their nervousness. Users were also a bit \textit{confused with terminology} used in the app and it made them \textit{uncomfortable}. Users often felt they weren't using the app correctly and couldn't provide good feedback [PU3]. This was also noticed by the developers. They felt bad seeing users confused and uncomfortable with the app [P5].

\begin{quote}
\small
    \faIcon{comments} \textit{``However, there was a point where I was little confused by the wording of something, so I felt then a little bit awkward, like I was not getting it right or not giving feedback...'' - PU3}
\end{quote}

\subsection{Strategies to Improve Empathy} \label{SEC:Strategies to Improve Empathy}
We describe the approaches developers used to support the users including \textit{technical support} and \textit{emotional support}. We also discuss prospective strategies they suggested when asked a hypothetical question on what would they do to better empathise with users in future. Prospective developer strategies include: \textit{Better preparation, Reduce the pace of the sessions, Ask more follow-up questions, Support when users are struggling, Taking time to implement the changes suggested by users, and More agile way of interacting with end users.} 

\subsubsection{Developer Support}
All developers explained different approaches for supporting users. They offered both technical and emotional support to users during these sessions. 

\textit{Technical Support.} During the usability sessions developers modified the questions based on the scenarios. They \textit{personalised the content}, rephrased and clarified questions, especially when users seemed confused [P1]. They stated that it was an attempt to support users to the best of their ability [P1]. 
\begin{quote}
\small
    \faIcon{comments} \textit{``I was able to try, I guess, modify the questions on the spot as, as the scenario changes, like throughout the meeting, to some extent. So I guess that maybe shows a bit of consideration towards each individual.'' - P1}
\end{quote}


Developers implicitly \textit{tried to talk in layman's terms} when dealing with the users [P3]. Developers were confident that the users sufficiently understood the technical details, which was evident in the way users performed tasks during the usability sessions. Users also confirmed this, stating that they were able to clearly understand and follow developer instructions. 
In addition, developers explicitly \textit{tried to phrase questions as clearly as possible at the first time} [P1, P6]. This demonstrates that the developers tried to support the users by giving the best possible instructions. 

\begin{quote}
\small
    \faIcon{comments} \textit{``We didn't really try to explain too much technical detail, at least in my sessions, we didn't really try to explain the technical details behind it unless it was like, strictly relevant. Often, it was more like,we want to do this, but we haven't, because we haven't had the time to implement it yet. Like that was kind of the most in depth...'' - P6}
\end{quote}

\textit{Emotional Support.} Developers described their \textit{urge to help users when they were struggling} during the usability sessions. However, they could not directly help the users as the goal of the sessions was to elicit unadulterated user feedback [P1]. Developers faced a dilemma of whether to empathise with users by helping them resolve their confusions, or allowing them struggle and find their own way. Developers were uncertain whether they found a balance in this dilemma. Despite feeling bad for confused users, developers often refrained from offering assistance. They believed they should let users resolve issues on their own to achieve the session's goal [P1, P5]. Based on some of the user responses, it was evident that occasionally developers have failed to find the balance in this situation [PU2, PU3]. Users said that they had requested developer assistance to clarify their confusions in such situations but it was not forthcoming. 

\begin{quote}
\small
    \faIcon{comments} \textit{``So I guess that's a scenario where I would empathise more with them when they're having struggles because I want to try to help them even and that's the is the point of the session to let them struggle to some extent so that we can get information about Yeah..'' - P1}
\end{quote}

When asked about a time where developers empathised with users, developers stated that they \textit{paid more attention to user concerns} when users brought up their own confusions [P1]. This suggests that supporting and wanting to support users is one of the things that comes to their mind when thinking of empathy.

\begin{quote}
\small
    \faIcon{comments} \textit{``Yes, I suppose like, as I said, given when the users gave feedback, so one of the users gave feedback on, like the wording was a bit confusing. And like when she pointed it out, it was really  easy for me to see that then. Like once you pay it once it was brought to it, and I paid more attention...'' - P2}
\end{quote}

Developers wanted to \textit{reassure users that they didn't need to strive for perfection} during usability sessions. They used phrases like "All good" to reassure users when they were worried about making mistakes, such as entering incorrect spellings or struggling to navigate the app. Some developers shared that they too had similar concerns, in order to reassure users in such situations [P2-P5].

\begin{quote}
\small
    \faIcon{comments} \textit{``Yeah, because obviously, in user testing, I don't want to say too much. But I also wanted to reassure [user] that there wasn't a way [user] could do it wrong...'' - P5}
\end{quote}


Developers \textit{reinforced the successful acts of users} to provide more support and they demonstrated it using the terms like “Perfect” [P6]. This suggests that empowering users by reassuring and reinforcing their successful acts is one of the innate empathetic behaviours of developers when directly interacting with users. This corroborates one of the verbal empathy cues which is affirming or showing understanding for the other person (Verbal empathy cue 4).

Developers described that their use of certain behaviours and words during usability sessions was driven by the intention to \textit{comfort users and connect} with them. They used phrases like "Perfect" or "All good" when users deviated from expected task performance and occasionally used humor too [P3-P5]. This indicates that developers naturally sought to create a rapport with users. They might have done this as a way of prompting for an effective user evaluation.

\begin{quote}
\small
    \faIcon{comments} \textit{``I think that's my main concern with that was like, trying to comfort her, so that you know, like don't think that you're doing anything wrong, there's nothing wrong with that..'' - P4}
\end{quote}

\subsubsection{Prospective Acts of Empathy}
Answering a hypothetical question on future acts of empathy, developers expressed they would work on \textit{phrasing their remarks in a better way} [P2] and \textit{spend additional time to prepare} and understand the tasks. Developers also acknowledged their nervousness during hosting the sessions and indicated their desire to prepare more to better engage users during these sessions for more productive outcomes [P1]. Additionally, they emphasised the importance of asking \textit{follow-up questions}, particularly concerning user feedback, to gain insights into user perspectives and thought processes [P1, P2]. This corroborates one of the verbal empathy cues which is asking a substantive question about the statement of the other person to better understand what the other person is saying (Verbal empathy cue 3).


\begin{quote}
\small
    \faIcon{comments} \textit{``...and like maybe ask follow-up questions. Especially like if a user has feedback like ask more questions about that, rather than just saying like, okay let's move on...'' - P2}
\end{quote}

Developers preferred \textit{a more agile approach to interact with users} as they wanted to know the real needs of users. They believed that this approach, which involves developing a working software prototype for usability sessions, would allow having consistent communication with users and the discovering their real needs. Developers highlighted the benefits of obtaining user feedback earlier in the development cycle and the potential for continuous feedback and improvement through this agile approach [P6].


\begin{quote}
\small
    \faIcon{comments} \textit{``All we do is we'd like to work much ourselves for a while for maybe like 7,8,9,10 weeks. And then we talk to the end users for like, two weeks, and then we'd go back and do your own thing. And then we talk to the end users again...but like having a more constant interaction with end users would make a more it would have affected both the developers and users. because we would have, as a developer, we would have had more constant need to discover what the end users want. And more constant need to develop prototypes and things that they can use, to show off in their sessions. And also, like, more room constant revisions...'' - P6}
\end{quote}


Developers also wanted to \textit{slow down the sessions} [P2] and \textit{take time to implement the changes suggested by users} [P6]. They preferred modifying the app incorporating user feedback which would allow users to see the implementation of their vision. Developers believed it would have been more beneficial to show the modified app to users during usability sessions as they are the ones who are really going to use it [P6]. 

\begin{quote}
\small
    \faIcon{comments} \textit{``we just wanted to make what we thought would be the best. And then listen to the end users and be like, that's great. We love your input. We'll change this thing. But like, we're still gonna go with what we want... 
    If we had taken a step and being like, Alright, now we know, we're gonna try to spend some time implementing the vision that the these end users had, because ultimately, they're the ones who are going to be using it, not us.'' - P6}
\end{quote}

Developers expressed their intention to \textit{provide better support to users when they are struggling} or confused. They acknowledged the need for careful preparation to manage unexpected user actions or difficulties, a consideration they had not thoroughly planned for previously [P5]. These strategies represent their commitment to improve empathy with end users in future sessions.


We posed the same hypothetical question to the users. The majority of users (three out of four) said they would not change anything as \textit{``everything was quite good and enjoyable''} [PU1, PU2, PU4]. One of the users stated that they would be more confident to provide feedback in future as they have experienced it once [PU3].

\subsection{Observed Empathy Cues (Observation Findings)} \label{SEC:Observation Findings} 

We identified the demonstration of several verbal and nonverbal empathy cues which are explained in Tables \ref{TAB:Verbal Empathy Cues} and \ref{TAB:Nonverbal Empathy Cues}. Below we discuss the observed empathy cues, focusing on how developers and users exhibited these cues and the number of occurrences of the cues.

\begin{table*} [ht]
\begin{threeparttable}
    \caption{Demonstration of Verbal Empathy Cues}
    \label{TAB:Demonstration of Verbal Empathy Cues}
    \footnotesize
    \begin{tabular}{lll}
        \toprule
        \textbf{Cue} & \textbf{Developer Demonstration} &  \textbf{User Demonstration} \\
        \midrule

        Empathic Understanding & \textit{H* \& MT*:} Answering user questions by repeating & Answering developers' questions by referring to\\
        Responses & \& referring to their previous content.  & their statements.\\
        & &\\

        Empathic Affirmations & \textit{H:} Summarising users' actions \& suggestions & Summarising and repeating developers' statements.\\
        & while responding to them. & Summarising developer content while explaining\\
        & Repeating users' statements. & user preferences.\\
        & \textit{MT:} Summarising users' suggestions \& struggles. &  \\
        & &\\

        Empathic Conjectures & \textit{H \& MT:} Asking follow-up questions from users to &  Asking follow-up questions from developers to\\
         & understand their responses \& suggestions. & understand the instructions.\\
        & &\\

        Empathic Evocations & \textit{H \& MT:} Praising users and showing & Praising developer effort and showing\\
        & understanding when users faced difficulties. & understanding.\\

        \bottomrule
        
    \end{tabular}

    \begin{tablenotes}
      \item \textit{\textsc{*} H: Hosting, MT: Minute Taking}
    \end{tablenotes}
    \end{threeparttable}
\end{table*}

\subsubsection{Verbal Empathy Cues}
We observed all four verbal cues (Table \ref{TAB:Verbal Empathy Cues}) from both developers and users. However, participants demonstrated these verbal cues differently, as shown in Table \ref{TAB:Demonstration of Verbal Empathy Cues}. When considering developers, we noticed that the majority of verbal cues were observed from the hosting developers. Among them, the most dominant verbal empathy cues were third (empathic conjectures: follow-up questions) and fourth cues (empathic evocations: praising other person). By demonstrating third verbal cue, hosting developers asked multiple follow-up questions to understand responses of users and suggestions provided by users. We noticed only a few follow-up questions from MT developers. 
Hosting developers exhibited the fourth empathy cue by praising users and showing understanding when users faced difficulties during usability testing sessions. 
The first cue was the least observed verbal empathy cue among hosting developers. Hosting developers displayed the first cue by repeating the content of users' statements while answering their questions. We hardly noticed verbal empathy cues from MT developers. We observed fourth verbal cue six times among them. When considering users the most observed verbal empathy cues were first and third cues. The least observed verbal empathy cue among users was the second cue where users summarised and repeated developers' statements. Occurrences of verbal empathy cues of developers and users are illustrated in Table \ref{TAB:Occurrences of Verbal Empathy Cues of Developers} and Table \ref{TAB:Occurrences of Verbal Empathy Cues of Users}. 

\begin{table}
    \begin{threeparttable}
    \footnotesize
    \parbox{.5\linewidth}{
    \centering
    \caption{Occurrences of Verbal Empathy Cues of Devs}
    \label{TAB:Occurrences of Verbal Empathy Cues of Developers}
    \begin{tabular}{ccccccccc}
        \toprule
        \textbf{Devs\textsc{*}} & \multicolumn{2}{c}{\textbf{V\textsc{*} Cue 1 \#}} & \multicolumn{2}{c}{\textbf{V Cue 2 \#}} & \multicolumn{2}{c}{\textbf{V Cue 3 \#}} & \multicolumn{2}{c}{\textbf{V Cue 4 \#}} \\

         & H\textsc{*} & MT\textsc{*} & H & MT & H & MT & H & MT \\
        \midrule

        P1 & 1 & 1 & 1 & 2 & 10 & 1 & 10 & 5\\
        P2 & 5 & 0 & 9 & 0 & 11 & 0 & 9 & 0\\
        P3 & 2 & 0 & 3 & 0 & 7 & 0 & 7 & 0\\
        P4 & 5 & 0 & 3 & 0 & 3 & 0 & 7 & 0\\
        P5 & 3 & 0 & 5 & 0 & 12 & 0 & 8 & 0\\
        P6 & 2 & 0 & 2 & 0 & 1 & 0 & 7 & 1\\
        \textbf{Total} & \textbf{18} & \textbf{1} & \textbf{23} & \textbf{2} & \textbf{44} & \textbf{1} & \textbf{48} & \textbf{6} \\
        \bottomrule    
    \end{tabular}
    \begin{tablenotes}
        \item \textit{\textsc{*} Devs: Developers, V: Verbal, NV: Nonverbal, H: Hosting, MT: Minute Taking}
    \end{tablenotes}
   }
   \hfill
    \parbox{.45\linewidth}{
    \centering
    \caption{Occurrences of Nonverbal Empathy Cues of Devs}
    \label{TAB:Occurrences of nonverbal Empathy Cues of Developers}
    \begin{tabular}{ccccccc}
        \toprule
        \textbf{Devs} & \multicolumn{2}{c}{\textbf{NV\textsc{*} Cue 1 \#}} & \multicolumn{2}{c}{\textbf{NV Cue 2 \#}} & \multicolumn{2}{c}{\textbf{NV Cue 3 \#}} \\

         & H & MT & H & MT & H & MT \\
        \midrule

        P1 & 23 & 9 & 63 & 98 & 8 & 34 \\
        P2 & 66 & 0 & 57 & 1 & 29 & 4 \\
        P3 & 22 & 0 & 40 & 21 & 11 & 9 \\
        P4 & 15 & 0 & 26 & 2 & 7 & 2 \\
        P5 & 13 & 1 & 14 & 12 & 3 & 4 \\
        P6 & 12 & 0 & 28 & 4 & 3 & 2 \\
        \textbf{Total} & \textbf{151} & \textbf{10} & \textbf{228} & \textbf{138} & \textbf{61} & \textbf{55} \\
        \bottomrule    
    \end{tabular}
    \begin{tablenotes}
        \item \textit{\textsc{*} Devs: Developers, V: Verbal, NV: Nonverbal, H: Hosting, MT: Minute Taking}
    \end{tablenotes}
   }
    \end{threeparttable}
\end{table}

\begin{table}[htbp]
    
    \parbox{.45\linewidth}{
    \centering
    \caption{Occurrences of Verbal Empathy Cues of Users}
    \label{TAB:Occurrences of Verbal Empathy Cues of Users}
    \footnotesize
     \begin{tabular}{ccccc}
        \toprule
        \textbf{Users} & \textbf{V Cue 1} & \textbf{V Cue 2} & \textbf{V Cue 3} & \textbf{V Cue 4} \\
        & \textbf{Count} & \textbf{Count} & \textbf{Count} & \textbf{Count} \\
        \midrule

        PU1 & 5 & 2 & 7 & 0 \\
        PU2 & 4 & 3 & 4 & 6 \\
        PU3 & 3 & 3 & 7 & 5 \\
        PU4 & 7 & 2 & 3 & 4 \\
        \textbf{Total} & \textbf{19} & \textbf{10} & \textbf{21} & \textbf{15}\\
       
        \bottomrule
        
    \end{tabular}
    \begin{tablenotes}
        \item \textit{\textsc{*} V: Verbal}
    \end{tablenotes}
   }
   \hfill
    \parbox{.5\linewidth}{
    \centering
    \caption{Occurrences of Nonverbal Empathy Cues of Users}
    \label{TAB:Occurrences of nonverbal Empathy Cues of Users}
    \footnotesize
    \begin{tabular}{cccc}
        \toprule
        \textbf{Users} & \textbf{NV Cue 1} & \textbf{NV Cue 2} & \textbf{NV Cue 3}  \\
        & \textbf{Count} & \textbf{Count} & \textbf{Count} \\
        
        \midrule

        PU1 & 29 & 13 & 13 \\
        PU2 & 47 & 21 & 11 \\
        PU3 & 11 & 21 & 28 \\
        PU4 & 29 & 10 & 21 \\
        \textbf{Total} & \textbf{116} & \textbf{65} & \textbf{73} \\
       
        \bottomrule
        
    \end{tabular}
    \begin{tablenotes}
        \item \textit{\textsc{*} NV: Nonverbal}
    \end{tablenotes}
   }
  
\end{table}

\subsubsection{Nonverbal Empathy Cues}
All nonverbal empathy cues (Table \ref{TAB:Nonverbal Empathy Cues}) were demonstrated by both developers and users during usability sessions. There were varied behaviours linked to each cue, as shown in Table \ref{TAB:Demonstration of nonverbal Empathy Cues}. The second nonverbal cue was the most observed cue among both hosting and MT developers. This cue was demonstrated by developers via facial expressions for attentive listening such as nodding, raised eyebrows, thumb \& forefinger of a hand on the chin and moving face closer to screen \cite{schmelzer2015empathie}. Third and first nonverbal cues were the least observed cues among hosting developers MT developers, respectively. Developers exhibited third and first nonverbal empathy cues respectively through their facial expressions similar to users and by verbally acknowledging user's speech via different sounds and short, monosyllabic words (see Table \ref{TAB:Demonstration of nonverbal Empathy Cues}). When considering users, first nonverbal cue was the most observed cue where as second cue was the least observed. Occurrences of nonverbal empathy cues of developers and users are illustrated in Table \ref{TAB:Occurrences of nonverbal Empathy Cues of Developers} and Table \ref{TAB:Occurrences of nonverbal Empathy Cues of Users}. 


       
        

\begin{table*} [htbp]
    \begin{threeparttable}
    \centering
    \caption{Demonstration of Nonverbal Empathy Cues}
    \label{TAB:Demonstration of nonverbal Empathy Cues}
    \footnotesize
    \begin{tabular}{lll}
        \toprule
        \textbf{Nonverbal Cue} & \textbf{Developer Demonstration} &  \textbf{User Demonstration} \\
        \midrule

        Lute for careful listening & \textit{H \& MT*:} Verbal acknowledgement via sounds like & Verbal acknowledgement via sounds like ``Mmm'', \\
         & ``Mmm'', ``Ah'' \& short, monosyllabic words such as & ``Ah'' \& short, monosyllabic words such as ``Ok'',\\
         & ``Ok'', ``Cool'', ``Yeah''. & ``Yeah''. \\
        & &\\

        Nonverbal signals for  & \textit{H \& MT:} Facial expressions for attentive listening & Facial expressions for attentive listening such as\\
        attentive listening & such as nods, raised eyebrows, thumb \& forefinger & nods \& raised eyebrows.\\
         & of one hand on the chin. & \\
        & Moving face closer to screen.&\\
        & &\\

        Similar facial expressions, & \textit{H \& MT:} Facial expressions similar to users such &  Facial expressions similar to developers such as\\
        posture, gestures & as laughing/smiling [Joy]. & laughing/smiling [Joy].\\
        
        \bottomrule
        
    \end{tabular}
        \begin{tablenotes}
            \item \textit{\textsc{*} H: Hosting, MT: Minute Taking}
        \end{tablenotes}
    \end{threeparttable}
\end{table*}

\subsection{Empathy Test Results} \label{SEC:Empathy Test Findings}
We calculated the empathy test scores of both developers and users in each round of development. In addition we calculated quartiles, interquartile range (IQR), maximum and minimum values of empathy test scores. These data were then used to draw box plots to demonstrate the overall dispersion of empathy scores of developers and users. 
We visualised empathy scores using several box plots. In this section we have only included the box plots drawn for total empathy, total cognitive empathy and total affective empathy scores of developers and users in each round. We explain the implications of these graphs based on the length and position of the box within and also across participant groups. 

\begin{figure}[htbp]
    \centering 
    \includegraphics[scale=0.5]{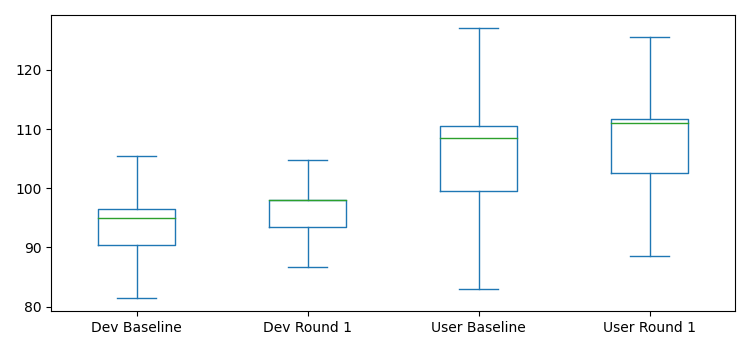}
    \caption{Total Empathy Distribution (Removed two outliers)}
    \label{FIG:Total empathy distribution}
\end{figure}




When considering the graphs that illustrate \textit{total empathy}, we noticed that developer box plots are short compared to user box plots (see Figure \ref{FIG:Total empathy distribution}) in both baseline and round one. This implies that overall developers have similar empathy scores whereas user empathy scores are quite diverse (Dev IQR < User IQR). It is clear that both developer and user box plots are shorter in round one compared to their baseline box plots. This implies that both developer and user total scores are less dispersed in round one compared to baseline empathy scores. 
The user box plots are much higher than developer box plots in both baseline and round one. This suggests that overall users have higher total empathy score compared to developers. The developer round one (Dev Round 1) box plot is much higher than baseline (Dev Baseline) box plot. This same behaviour can be seen with respect to the box plots of user baseline and round one empathy. This suggests that overall developer and user empathy has increased in round one compared to their baseline empathy. 

We noticed two outlier scores for a developer and a user in both baseline and round one empathy who scored below the lower whisker\footnote{The whiskers are the two lines outside the box, that go from the minimum to the first quartile (lower whisker) and then from the third quartile to the maximum (upper whisker).} limit. During the interview, this developer was uncertain whether there's any need for empathising with users in this project. It was evident from this developer's statement \textit{``I don't really think, I don't know if there was much to empathise with the customers in this particular project''}.
However, we do not have a deeper understanding regarding the user with the outlier score. Based on our experience in interviewing this user, we assume this user may not have understood the scale items properly as these scale items were very generic. 

\begin{figure}%
\begin{minipage}[t][3.7cm][t]{0.485\textwidth}
        \includegraphics[width=\textwidth]{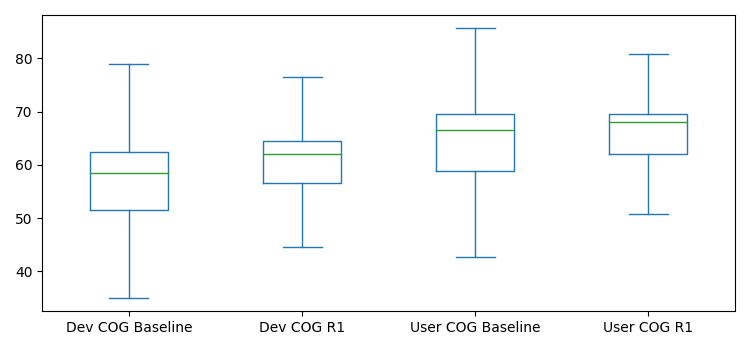}
        \caption{Total Cognitive Empathy Distribution (Removed an outlier)}
        \label{FIG:Cognitive empathy distribution}
    \end{minipage}
    \hfill
    \begin{minipage}[t][3.7cm][t]{0.485\textwidth}
        \includegraphics[width=\textwidth]{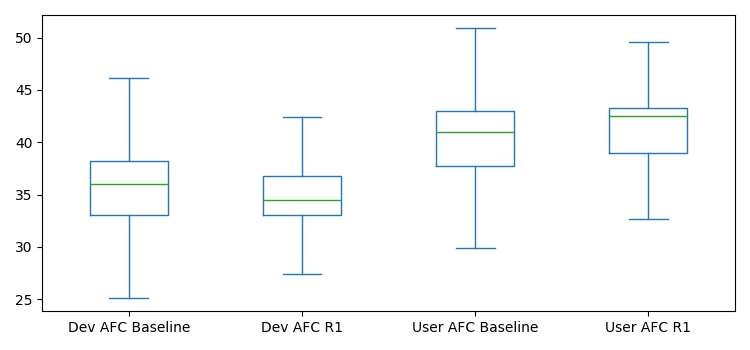}
        \caption{Total Affective Empathy Distribution (Removed an outlier)}
        \label{FIG:Affective empathy distribution}
    \end{minipage}%
\end{figure}

When considering the \textit{total cognitive empathy graphs} (Figure \ref{FIG:Cognitive empathy distribution}), it is clear that the user box plots are slightly shorter compared to the developer box plots (User IQR < Dev. IQR) in both baseline and round one. This implies that cognitive empathy scores of developers are more dispersed compared to users' cognitive scores. Both developer and user box plots are shorter in round one compared to their baseline box plots. This implies that both developer and user cognitive scores are less dispersed in round one compared to baseline empathy scores. 
User box plots are much higher than developer box plots in both baseline and round one. This suggests that overall users have higher cognitive empathy score compared to developers. The developer round one box plot is comparatively higher than developer baseline box plot. This infers that overall developer cognitive empathy scores are higher in round one compared to baseline scores. We noticed cognitive empathy box plot of users in round one is slightly higher than their baseline box plot. 
This implies that overall user cognitive empathy scores have increased in round one compared to their baseline scores.

We noticed an outlier score for a user in round one cognitive empathy who scored below the lower whisker limit. However, we noticed that user's cognitive score in round one has increased compared to baseline score. Despite this situation, user's score was identified as an outlier. This suggests that even though user's cognitive empathy score increased in round one, it has not increased much compared to other users. 

When considering the \textit{total affective empathy graphs} (Figure \ref{FIG:Affective empathy distribution}), both developer and user box plots are equal in length in baseline (Dev. IQR = User IQR). Despite this, the developer box plot became slightly shorter compared to user box plot in round one (Dev. IQR < User IQR). This illustrates that affective empathy scores of users are much dispersed compared to developers' in round one. Both developer and user box plots in round one are shorter compared to their baseline box plots. This implies that both developer and user affective scores are less dispersed in round one compared to baseline empathy scores. 
User box plots are much higher than developer box plots in both baseline and round one. This implies that overall users have higher affective empathy score compared to developers. We noticed that developer box plot in round one is comparatively lower than baseline empathy box plot. Even though the value of first quartile (Q1) is same and minimum is increased, all the other values are decreased (Q2, Q3, maximum) in round one compared baseline. This implies that overall affective empathy of developers has decreased in round one compared to baseline. However, we noticed that users' round one affective empathy box plot is higher than baseline box plot. This implies that users' overall affective empathy has increased in round one compared to baseline. 
Despite this trend we noticed an outlier score for a user in round one affective empathy who scored below the lower whisker limit. We identified that this user's affective empathy has reduced even when considering the numerical scores. However, we are not clear what could be the reason for this behaviour. 

\section{Discussion} \label{SEC:Discussion}

\subsection{Collating Mixed Methods Data}
Qualitative measures, including interviews and observations, were instrumental in uncovering empathy enablers, barriers, and strategies to overcome barriers. These qualitative methods also revealed the manifestation of empathy cues during direct interactions between developers and users, offering insights into how empathy is practised based on their experiences and interactions. Complementing these qualitative approaches, we employed quantitative measures, specifically empathy test scores determined by the QCAE empathy scale. These scores offered a numerical assessment of participants' empathetic tendencies, providing a comprehensive understanding of empathy in our study.

Our demographics questionnaire helped us to understand our participants better and it helped us to be more open during our observations. We observed empathy cues and behaviours of participants during usability sessions. These observations informed the interviews and we were able to inquire about different behaviours of participants. Empathy test scores provided a quantitative measure to understand participants' empathy. This measure also informed our interviews by helping us to compare \& contrast participants' scores against demonstrated empathy cues and explanations provided during interviews. Hence our observations and empathy test scores drove our interviews in a better direction by helping us to form better questions and capture rich data.

We found several correlations between empathy cues, interview findings and empathy scores. We were able to map third and fourth verbal empathy cues (Table \ref{TAB:Verbal Empathy Cues}) with the enablers and strategies identified in interview findings. The fourth verbal cue is associated with praising or showing understanding for the other person. It is rooted in the codes \textit{``Empathy due to appreciation''} and \textit{``Empathy towards user confusion''}. These codes belong to the enablers of empathy and usage of mixed methods assisted in unfolding the connection between empathy cues \& enablers of empathy. We also identified that the fourth verbal cue is embedded in the developer strategies such as \textit{``Reassuring users''} and \textit{``Reinforcing the successful acts of users''}. Developers used these strategies to emotionally support users. The third verbal cue which is associated with substantive questions to better understand the other person is rooted in the codes \textit{``Developer plans to prepare more''} and \textit{``Developer plans to ask more follow-up questions''}. These codes belong to the prospective developer acts of empathy which comes under the developer strategies. 

We identified correlations between empathy cues and empathy scores. While individual empathy scores for developers and users were not analysed due to insufficient data, we opted to assess the scores across two groups: developers and users. This approach allowed us to measure changes in empathy within each group over time, providing valuable insights into the general empathetic dynamics between developers and users throughout the study period.
When considering developers, we noticed that the highest number of empathy cues were reported from the hosting developers. We suspect this was due to the less involvement of MT developers. Most of the developers stated that they thought they were not supposed to talk with users while taking minutes (P3, P4, P5). We rarely noticed any communication between MT developers and users while observing the usability sessions. Hence verbal empathy cues were by default low in MT role. This could be a reason for low number of overall empathy cues reported from developers while performing MT role compared to hosting. However, the highest number of nonverbal cues were also observed from hosting developers. Several developers stated that they had a minimal interaction with users while performing MT role and it was a barrier to their empathy (P3, P4, P5). This could be a reason for low number of nonverbal empathy cues. 
We noticed some patterns between overall empathy cue counts and empathy test scores of participants in round one \& round two. We identified three main patterns: high empathy cue count when empathy score is high; low empathy cue count when empathy score is high; and having the same empathy scores. We had three instances where participants had the same or almost the same empathy scores in all rounds. These participants were developers and in all these instances we noticed that the empathy cue count is high when hosting the session. We also identified four participants who had high empathy cue count when total empathy score is high. Two of these participants were developers and they both demonstrated high empathy cue count when hosting. We identified two participants who had low empathy cue count when empathy is high. One of them was a developer and this developer demonstrated this high empathy score while performing MT role. We also had one user who participated only in the first round which limits our ability to compare this user's empathy score with empathy cue count. Most of the time we noticed developers had a high empathy cue count when they were hosting (five out of six). Sometimes their empathy cue counts did not reflect their empathy score. However, their cue counts were more representative of the role they played in the usability testing sessions. This could be due to the nature of usability sessions.

\subsection{Cognitive \& Affective Empathy}
Most of the empathy enablers that we identified in this study can be categorised as cognitive and affective. Cognitive empathy enablers triggered cognitive empathy. Likewise affective empathy enablers triggered affective empathy. Enablers such as \textit{reflection in action, familiarity, improved user understanding, user confusions, user interaction, similar levels of technical ability, understanding of developer thought process and understanding of developer needs} were identified as cognitive empathy enablers. These helped participants to detect and understand others' perspectives. For instance, familiarity was a cognitive empathy enabler for developers. In this context developer familiarity refers to closeness to code and familiarity of user issues. This familiarity helped developers to better understand user suggestions and user issues, and is a demonstration of enabling cognitive empathy. 

Enablers such as \textit{awareness of user actions and emotions, appreciation, technology literacy related issues, gender, familiarity of developers \& connection, connection to the app's goal and appreciation of developer efforts} were identified as affective empathy enablers. These enablers triggered participants to react to others' emotions and share an emotional experience. For instance, familiarity of developers \& connection was an affective empathy enabler for users. This familiarity \& connection made users feel emotionally close to developers, and users were able to establish a rapport with developers, demonstrating enabling of affective empathy.

\subsection{Empathy Enablers and Barriers} \label{SEC:Empathy Guide}
As illustrated in Figure \ref{FIG:Interview Findings}, our STGT data analysis uncovered various developer awareness types (see Table \ref{TAB:Types of Developer Awareness}), empathy enablers (see Table \ref{TAB:Enablers of Developer Empathy}), barriers to developer and user empathy (see barriers in Table \ref{TAB:Developer Empathy Barriers and Strategies} and Table \ref{TAB:User Empathy Barriers and Strategies}), and strategies for overcoming these barriers (see Table \ref{TAB:Developer Strategies}).
These identified enablers and barriers align with established empathy literature, particularly drawing parallels with enablers such as familiarity and commonality \cite{motomura2015interaction, thompson2019empathy, batson1996prior, eklund2009similar, heinke2009cultural, hoffman2001empathy, yaghoubi2023young, preis2012pain, xu2009pain, yaghoubi2021histories,snow2000empathy}  and barriers including time pressure, disinterest, and task-centeredness \cite{howick2017barriers, taleghani2018barriers}.
We identified that developer awareness of their own emotions \& actions, as well as emotional \& technical awareness of users, is a first step of triggering empathy. This awareness may assist in enabling empathy. Goleman also endorsed the notion that empathy builds on self-awareness  \cite{goleman1996emotionalintelligence}. If empathy is not triggered, then this awareness most likely leads developers to identify barriers to empathy. When developers identify these empathy barriers, they may apply some strategies to overcome these barriers. 

Developer awareness reinforced enablers of empathy and developers used some strategies to overcome their empathy barriers. For instance, developers had the technical awareness of the users' understanding. This improved understanding enabled developers' empathy towards users' technology literacy related issues (see Table \ref{TAB:Types of Developer Awareness}, \ref{TAB:Developer Awareness and Enablers of Empathy}). However, some developers identified the difficulty in empathising with low-technology literacy related user concerns as a barrier to their empathy. To overcome this empathy barrier, developers are planning to follow a set of strategies in future including ask more follow-up questions, pay more attention to user concerns, and connect more with users (see Table \ref{TAB:Developer Strategies}, \ref{TAB:Developer Empathy Barriers and Strategies}).

Table \ref{TAB:Developer Awareness and Enablers of Empathy} provides examples of required developer awareness and empathy enablers while Table \ref{TAB:Developer Empathy Barriers and Strategies} presents developers' empathy barriers and strategies to overcome these barriers. We established these connections based on the analysis of data obtained through the interviews. Participants were asked about their experiences with empathy enablers, empathy barriers, the actions they took to overcome these challenges, and their proposed strategies for addressing such barriers in the future. Our STGT data analysis of these responses allowed us to pinpoint specific enablers for each type of awareness, and specific barriers and their associated strategies. We present these details exclusively for developers as our data is not sufficient to compile a similar guide for users. 
We propose some potential strategies developers can follow that may minimise barriers to users' empathy in Table \ref{TAB:User Empathy Barriers and Strategies}. For instance, users highlighted confusion on terminology as a barrier to empathy. When developers observe users are confused about certain terminology, we propose developers to employ strategies such as rephrasing \& clarifying, using simple language and avoiding overloading users with technical details. 

\begin{table}[htbp]
    
    \parbox{.5\linewidth}{
        \caption{Types of Developer Awareness}
        \label{TAB:Types of Developer Awareness}
        \setlength{\aboverulesep}{0pt}
        \setlength{\belowrulesep}{0pt}
        \setlength{\extrarowheight}{.3ex}
        \footnotesize
        \begin{tabular}{lll}%
            
            \toprule
            \textbf{ID} & \textbf{Type} &  \textbf{Nature of Awareness} \\
            \midrule
    
            \rowcolor{tableGray}
            A1 & Self & Developer awareness of technical jargon \\
            
            \rowcolor{tableGray}
            & Awareness & Developer awareness of differences in empathy \\

           A2 & Emotional & Users being stressed due to recording\\
            
            & Awareness & Users feeling pressured\\
            & & Developers noticing user reactions\\
            & & Users might have felt developers sounded judgmental \\
            & & Developer awareness of user actions and emotions \\
            & & Developer understanding on user concerns \\
            & & Improved developer understanding on user needs \\

            \rowcolor{tableGray}
            A3 & Technical & Developer awareness of the importance of empathy \\
            \rowcolor{tableGray}
            & Awareness & towards users\\
            
            \rowcolor{tableGray}
            & & Developers' awareness of user's technical proficiency\\
            
            \rowcolor{tableGray}
            & & Developer awareness of user understanding\\
    
    
            \bottomrule
        \end{tabular}%
    }
    \hfill
    \parbox{.45\linewidth}{
        \caption{Enablers of Developer Empathy}
        \label{TAB:Enablers of Developer Empathy}
        \footnotesize
         \begin{tabular}{ll}%
            \toprule
            \textbf{ID} & \textbf{Enablers of Developer Empathy}\\
            \midrule
    
            E1 & Empathy towards user confusion\\
            E2 & Empathy due to commonality\\
            E3 & Empathy due to familiarity \\
            E4 & Empathy due to the awareness of user actions and emotions\\
            E5 & Empathy due to appreciation\\
            E6 & Empathy due to more developer involvement\\
            E7 & Empathy due to gender similarities\\ 
            E8 & Empathy due to better connection\\
            E9 & Empathy due to user interaction\\
            E10 & Empathy due to improved user understanding\\
            E11 & Empathy for technology literacy related issues\\
            E12 & Empathy due to similar levels of technical ability\\
            E13 & Empathy due to reflection in action\\
            \bottomrule
         \end{tabular}%
    }
\end{table}%

\begin{table*} [h]
    \centering
    \caption{Developer Awareness and Enablers of Empathy}
    \label{TAB:Developer Awareness and Enablers of Empathy}
    \footnotesize
    \begin{tabular}{lccccccccccccc}
         \toprule
        \textbf{Type of Awareness} & \multicolumn{12}{c}{\textbf{Enablers of Developer Empathy}}\\
         \midrule
         & E1 & E2 & E3 & E4 & E5 & E6 & E7 & E8 & E9 & E10 & E11 & E12 & E13\\
          
          \rowcolor{tableGray}
          Self Awareness &  \faCheckSquare & &  &  & & & & & & & & & \\ 
         
          Emotional Awareness &  & \faCheckSquare & \faCheckSquare & \faCheckSquare & \faCheckSquare & \faCheckSquare & \faCheckSquare & \faCheckSquare & \faCheckSquare & & & & \\ 

          \rowcolor{tableGray}
          Technical Awareness & \faCheckSquare & \faCheckSquare & \faCheckSquare & & & & & & \faCheckSquare & \faCheckSquare & \faCheckSquare & \faCheckSquare & \faCheckSquare\\ 
          
        \bottomrule
    \end{tabular}
\end{table*}

\begin{table}[htbp]
        \caption{Developer Strategies}
        \label{TAB:Developer Strategies}
        \footnotesize
         \begin{tabular}{llll}%
            \toprule
            \textbf{ID} & \textbf{Strategies to Overcome Barriers} & \textbf{ID} & \textbf{Strategies to Overcome Barriers}\\
            \midrule
    
            S1 & Developer plans to prepare more & S10 & Developer thoughts to better phrase the remarks\\
            S2 & Developer attempts to connect more with users & S11 & Reassure users\\
            S3 & Provide better support when users are struggling & S12 & Rephrase and clarify when users seemed confused\\
            S4 & Pay more attention to user concerns & S13 & Developer attempt to talk in layman’s terms\\
            S5 & Developer plans to slow down the session & S14 & Not overloading users with technical details\\
            S6 & Developer plans to ask more follow-up questions & S15 & Reinforcing the successful acts of users\\
            S7 & Developers having the urge to help users when they struggle to use the app & S16 & Taking time to implement user suggested changes\\
            
            S8 & More agile way of interacting with end users & S17 & Think thoroughly before talking\\
            S9 & Developer attempts to make users more comfortable & S18 & Personalising content based on the users\\
            \bottomrule
         \end{tabular}%
\end{table}%

\begin{table*} [htbp]
    \caption{Developer Empathy Barriers and Strategies}
    \label{TAB:Developer Empathy Barriers and Strategies}
    
    \footnotesize
    \begin{tabular}{lccccccccc}

        \toprule
        \textbf{Developer Empathy Barriers} & \multicolumn{9}{c}{\textbf{Strategies to Overcome Barriers}}\\
        \midrule
        & S1 & S2 & S3 & S4 & S5 & S6 & S7 & S8 & S9 \\

        \rowcolor{tableGray}
        Less developer interest & \faCheckSquare & \faCheckSquare & & & & & & & \\
        
        Task centredness & & \faCheckSquare & \faCheckSquare & \faCheckSquare & & & & & \\

        \rowcolor{tableGray}
        Unfamiliarity of features & \faCheckSquare &  & & & & & & & \\

        Limited time for reflection in action & & & & & \faCheckSquare & \faCheckSquare & & & \\

        \rowcolor{tableGray}
        Difficulty in understanding user struggles & & \faCheckSquare & & \faCheckSquare  & & \faCheckSquare & \faCheckSquare & & \\

        Difficulty in empathising with low-tech. literacy related issues & & \faCheckSquare & & \faCheckSquare & & \faCheckSquare & \faCheckSquare & & \\

        \rowcolor{tableGray}
        Less interaction & & \faCheckSquare & & & & & & \faCheckSquare & \faCheckSquare  \\

        Difficulty to resonate with collective experiences of women & \faCheckSquare & \faCheckSquare & & & & & & & \\
       
        \bottomrule
    \end{tabular}

\end{table*}

\begin{table*} [htbp]
    \caption{User Empathy Barriers and Strategies}
    \label{TAB:User Empathy Barriers and Strategies}
    \footnotesize
    \begin{tabular}{lccccccccccccc}
        
        \toprule
        \textbf{User Empathy Barriers} &  \multicolumn{13}{c}{\textbf{Strategies to Overcome Barriers}} \\
        \midrule
        & S2 & S4 & S7 & S9 & S10 & S11 & S12 & S13 & S14 & S15 & S16 & S17 & S18\\

        \rowcolor{tableGray}
        Poor user connection with developers & \faCheckSquare & & & & & & & & & & & &  \\

        User nervousness due to time confusions & & & & \faCheckSquare & & & & & & & & &  \\

         \rowcolor{tableGray}
         Users being nervous by thinking about session questions & & & & \faCheckSquare & \faCheckSquare & \faCheckSquare & & & & & & &  \\

         User nervousness due to unfamiliarity of sessions & & & & & \faCheckSquare & \faCheckSquare & \faCheckSquare & \faCheckSquare & \faCheckSquare & \faCheckSquare & & &  \\

         \rowcolor{tableGray}
         User nervousness about providing useful feedback & & & & & & \faCheckSquare & & & & \faCheckSquare & \faCheckSquare & &  \\

         User confusion on terminology & & & & & & \faCheckSquare & \faCheckSquare & \faCheckSquare & \faCheckSquare & & & \faCheckSquare & \faCheckSquare \\
         
         \rowcolor{tableGray} 
         Users feeling awkward due to confusion & & \faCheckSquare & \faCheckSquare & \faCheckSquare & & \faCheckSquare & & & & & & &  \\ 
         
    \bottomrule
    \end{tabular}
\end{table*}

\subsection{Key Insights \& Reflections} \label{SEC:Insights}
We determined the following insights, reflections and emerging relationships based on our data analysis.

\faIcon{cogs} \textbf{Grateful users:} We noticed the grateful nature of the users. Users expressed their gratitude in many forms, both during the usability sessions directly to developers and in the interviews. It seems that the context of the application (eHealth) is one of the factors which influenced this grateful behaviour. Users were very grateful to developers for building an application which would assist them as PCOS patients. During the usability sessions users commended the developers for bringing their vision into life. Users seemed very pleased by the overall application and available features, and they even provided constructive feedback, suggesting improvements and new features like mood tracking. Notably, users did not provide negative feedback and expressed their intention to use the AskPCOS app when it is released. Users appreciated the opportunity of participating in the usability sessions and commended the developers' flexibility. Even during the interviews, users expressed a positive sentiment about their overall experience.

\faIcon{cogs} \textbf{Reflection in action:} Reflection in action helped developers to empathise and connect with the users. However, the limited time for reflecting during the sessions \textit{(reflection in action)} was a barrier to developer empathy. In addition, developers had difficulties in understanding user feedback due to reflection on action. This situation aligns with the idea of Sch{\"o}n's reflective model \cite{schon2017reflective} in professional education, emphasising the benefits of reflection in action (during event) over reflection on action (post-event). 
Our findings suggest that reflection in action enables developer empathy and reflection on action hinders developer empathy.

\faIcon{cogs} \textbf{Closeness to code:} We identified developer familiarity has two parts: closeness to code and familiarity with user issues. Software developers are known to be defensive about their code and technical approaches \cite{mcavoy2009failure}. However, we found that closeness to code enabled developer empathy. Developers showed more empathy towards user suggestions concerning the features they were closely working on. 
This familiarity helped developers to better understand user suggestions and empathise with their experiences. 

\faIcon{cogs} \textbf{Connection Vs familiarity Vs empathy:} We noticed an emerging relationship among familiarity, connection and enablers of empathy. We found that familiarity helped to establish a better connection and having a better connection enabled empathy. We also found that familiarity triggered empathy. 
Further we found that unfamiliarity led to a poor connection \& it was also a barrier to empathy. Although it is said that \textit{`familiarity breeds contempt'}, we noticed that \textit{familiarity breeds empathy}. We noticed these emerging relationships from both developers and users, suggesting a correlation among connection, familiarity, and empathy.
Our insights align with established empathy research, supporting the notion that familiarity serves as an enabler for empathy \cite{motomura2015interaction}. Further, the framework developed by Thompson et al. emphasises that the individuals who are psychologically closer, and thereby more familiar, may evoke heightened empathetic responses \cite{thompson2019empathy}. This alignment with existing research strengthens our understanding of the interplay between connection, familiarity, and empathy in the dynamic context of our study.

\faIcon{cogs} \textbf{Interaction Vs connection Vs empathy:} We observed an emerging relationship between interaction, connection and enablers of empathy. We found that interaction led to a better connection and a better connection enabled empathy. We also found that interaction enabled empathy. We identified the less interaction as a barrier to developer empathy and users also identified less interaction as a cause of poor connection with developers. These insights suggest that there could be a correlation between interaction, connection, and empathy.


\faIcon{cogs} \textbf{Relatedness trigger empathy:} Our observations align with existing empathy research, which suggests that shared experiences foster empathy (Section \ref{SEC:Related Works}). Various studies corroborate that relatedness acts as an enabler for empathic behaviour \cite{batson1996prior, eklund2009similar, heinke2009cultural, hoffman2001empathy, yaghoubi2023young, preis2012pain, xu2009pain, yaghoubi2021histories, snow2000empathy}. We saw evidence of this on the aspects of technology and gender. In this context, developers' technical competence and their ability to empathise with users' technical difficulties, as well as shared gender experiences, facilitated better understanding and empathy toward users. This substantiates the notion that relatedness, whether in technical expertise or gender, helps in enabling empathy within the context of our study.


\faIcon{cogs} \textbf{Emotions and empathy:} We observed an emerging relationship between emotions and empathy. Emotions play a significant role in enabling or inhibiting empathy, as both positive and negative emotions were linked to empathy in our study. Positive emotions seemed to enable empathy, while negative emotions could either trigger or hinder empathy depending on the context and individual experiences. For instance, positive emotions like \textit{feeling fulfilled} and \textit{feeling understood} enabled developers' and users' empathy respectively. In addition, even negative emotions like \textit{guilt} triggered developer empathy. However, negative emotions like developer \textit{frustration} and user \textit{nervousness} were barriers to their empathy. 
Building upon existing research, we draw upon an integrative framework on empathy and emotion regulation by Thompson et al. to contextualise our findings \cite{thompson2019empathy}. This framework views empathy as a cybernetic, emotion-generative process and emphasises the role of regulatory strategies. Regulatory strategies like situation selection and cognitive change were found to influence distinct components of empathy and the resulting affective state in observers. Our study contributes to this theoretical framework by providing empirical evidence of how emotions, both positive and negative, interact with empathy in the dynamic context of software development.

\subsection{Possible Implications for Software Practitioners}

\faIcon{laptop} \textbf{Recommendation 1 -- Proactive identification of empathy enablers and barriers:} We have suggested some actions (Section \ref{SEC:Empathy Guide}) that might be considered by practitioners to identify the awareness required to enable empathy, determine enablers \& barriers to empathy, and some strategies to overcome these barriers. Although there is no clear understanding on the impact of empathy on software engineering per se, empathy is highly beneficial for improving human connections. Having a healthy connection among software practitioners, users and even customers may positively influence the success of software projects. 

\faIcon{laptop} \textbf{Recommendation 2 -- Facilitate direct user feedback to developers:} We noticed developers were very satisfied to receive positive user feedback. Positive feedback boosted developer confidence in their product and made them feel fulfilled. Direct interaction also helped developers better understand user pain points. Receiving feedback directly from end users may have had a more significant impact compared to the feedback developers typically receive from their managers. Thus, conducting usability testing sessions with developers and end users before launching software would be beneficial in fostering developer empathy towards user needs.



\faIcon{laptop} \textbf{Recommendation 3 -- Pilot sessions to find best suited developer roles:} We found that different developers preferred different session roles and their ability to interact with the users depended on their meeting role to some extent. Hence, we believe tailoring the session roles to the preferences and strengths of individual developers could have enhanced their ability to interact with users and understand feedback. In a real world setting, it would be beneficial to conduct a pilot session to identify the best suited role for each developer.


\faIcon{laptop} \textbf{Recommendation 4 -- Find ways to support struggling users without compromising their feedback:} We identified that developers were in a dilemma in wanting to empathise \& help users, and their need to elicit unadulterated user responses to usability concerns. This created a trade-off between helping users and letting them struggle. 
There could be a better approach where developers can provide necessary emotional support to users without providing direct technical support, thus balancing empathy and feedback authenticity. 

\subsection{Implications for Researchers}

\faIcon{graduation-cap} \textbf{Designing human experiments for empathy research:} We identified several factors that made both developers and users uncomfortable. We believe these could be better managed when designing empathy research experiments involving humans. First, we determined that developers preferred different meeting roles and their ability to interact with the users depended on these roles. Having an understanding on the preferences and capacity of each participant may assist in designing a better experiment. We noticed a developer dilemma in helping users and having proper strategies to manage such difficult situations would have been more beneficial. Hence we recommend considering these aspects when designing similar experiments. 

\faIcon{graduation-cap} \textbf{Validate with a large sample:} A key contribution of our study is a set of actions (Section \ref{SEC:Empathy Guide}) which can be followed to build awareness as well as to identify empathy enablers, barriers and strategies to deal with these barriers. However, our findings are based on a limited number of participants and all of our developers were students. Hence researchers may consider replicating our study with more participants and including experienced software practitioners. This new study will help to validate and strengthen our findings. 

\faIcon{graduation-cap} \textbf{Impact of empathy on SE and an SE-oriented empathy scale:} Our study did not focus on the impact of empathy on SE. It would be interesting to know the impact of empathy on software itself as well as software projects, software teams, and even other stakeholders such as customers, users, collaborators. 
Similarly, while designing and executing our study, we identified that it is quite difficult to apply existing empathy scales, mainly sourced from psychology. An SE-oriented empathy scale with SE-specific empathy measures or adapting existing scales to SE would be beneficial \cite{gunatilake2023empathy}. 

\faIcon{graduation-cap} \textbf{Developer awareness of user needs and emotions:} We found that developers were very attentive to users’ emotional and technical needs. Direct interaction between developers and users could be a reason for this developer awareness. Developers being well aware of users’ emotional and technical needs likely improved the usability of the apps.
Developers noticed both positive and negative user emotions, and were more concerned about negative emotions. It is unclear how much of this behaviour is \textit{Emotion Contagion} or \textit{Perspective Taking} or \textit{Empathic Concern} or is it \textit{Social Awareness} or is it \textit{Sympathy} or else is it just trying to be nice to users as they have a sensitive health condition. Developers were considerate about how users were feeling not just during the usability sessions, but even during interviews when reflecting on the usability sessions.
Studying the impact of increasing direct interactions between developers and users may be beneficial to better understand this awareness of user needs and emotions.

\faIcon{graduation-cap} \textbf{Developer connection with users and grateful users:} During interviews, developers occasionally referred to empathy as connection. We noticed evidence of a relationship between empathy and connection (Section \ref{SEC:Insights}). Unfamiliarity and less interaction were the major causes of poor developer connection, and these were also identified as barriers to developer empathy. This suggests a correlation between the connection and empathy. 
Further, users expressed gratitude towards developers and none of the users had negative feedback regarding the application (Section \ref{SEC:Insights}). We hypothesise this may be a common behaviour among eHealth app users. Future studies could explore the relationship between empathy and connection and user gratitude in the context of eHealth apps.

\faIcon{graduation-cap} \textbf{Reflection in action and closeness to code:} Findings suggest that reflection in action enables developer empathy, and reflection on action hinders developer empathy (Section \ref{SEC:Insights}). We also identified closeness to code as an empathy enabler. However, due to limited number of participants, we cannot generalise these findings. 


\faIcon{graduation-cap} \textbf{Emerging relationships:} We observed an emerging relationship between familiarity, connection and empathy as well as between interaction, connection and empathy. In addition we noted emerging relationships between relatedness \& empathy, and emotions \& empathy. These emerging relationships are worth studying further. 

\faIcon{graduation-cap} \textbf{Varying effects of empathy enablers and barriers over time:} While our primary means of identifying empathy enablers and barriers relied on semi-structured interviews, it is important to note that these interviews took place at the conclusion of the 24-week study period. Therefore, we faced limitations in capturing the dynamic changes and evolving effects of enablers and barriers over time. Additionally, other contextual factors influencing these dynamics could not be explored due to the retrospective nature of our interviews. Future research should consider longitudinal data collection methods to better understand the temporal aspects of empathy dynamics in software development contexts.

\section{Limitations} \label{SEC:Threats to Validity}
In our study, data collection was limited to a specific context, with participants exclusively composed of final year undergraduate students majoring in IT. While our sample size included a substantial number of developers, it is crucial to acknowledge the limited diversity in our participant pool, as they were predominantly from Australia. Consequently, the findings derived from our study may not be universally applicable to the broader community of software practitioners and users.
Given that the majority of our participants were students in their final year of undergraduate studies in IT from Australia, the external validity of our results is constrained, and the generalisability of our case study is thereby limited. It is important to recognise that our research scope focused on a particular demographic within the SE community, which may not fully represent the diverse landscape of practitioners worldwide.
To enhance the robustness of our conclusions, future studies should consider replicating this research with a more heterogeneous sample. Including experienced software practitioners from various professional backgrounds, as well as expanding the study to different countries and contextual settings, would contribute to a more comprehensive understanding of the phenomena under investigation. In addition, deeper statistical analysis could be done on future larger datasets. This approach would mitigate the limitations associated with external validity and provide a more nuanced and widely applicable insight into the broader SE community. 
Further the implications presented for software practitioners need to be validated through additional studies, as they are derived from this specific case study.
 
Although there are many empathy tests, none of the scales were designed specifically for SE context. We selected QCAE with the recommendation of empathy experts as it seems the best matching scale to our study. 
However, participants' understanding on the empathy test can vary. Different human aspects such as age, gender, ethnicity, and other psychological differences may influence the ratings of items on the QCAE. 
The self-report nature of the scale may have an impact on the overall score, thus participant empathy scores may be less reflective of how they actually demonstrated empathy. Empathy is regarded as a favourable trait and participants may have been tempted to respond in a more socially desirable way while filling the empathy test as well as during the interviews. We considered only baseline and round one empathy scores of developers and users. Even though we have round two scores of all the developers, we do not have round two score of one user. Hence we are using only baseline and round one scores while reporting our findings to make the presentation of data similar among both groups of participants. 

Usability testing sessions were conducted using Zoom platform due to the COVID-19 pandemic and geographical distribution of the users. Therefore our observation study was also carried out online. We may have missed some empathy cues due to the limitations in Zoom setting while observing participants. We tried to rectify this situation by having two observers, video recording these sessions and watching these recordings multiple times. However, due to human errors we may still have missed some of the empathy cues. 
In the interviews, developers pointed out that users might have felt stressed or under pressure due to the recording of the usability sessions. This could have potentially limited users' capacity to express themselves openly, representing a limitation of the study. The follow-up interviews allowed participants to retrospectively assess the conduct of the usability sessions and suggest improvements. Incorporating a similar approach earlier in the study could have been beneficial.

We used STGT for qualitative data analysis. We generated concepts, subcategories and categories based on the codes. However, codes generated by a single researcher could be subjective and can lead to a potentially limited view of data. Hence after the first author conducted the initial coding and analysis, it was shared with all the other authors to resolve any conflicts. The third author who is well-experienced in STGT, peer reviewed all the codes, concepts, subcategories and categories. We discussed all the conflicts and carried out several discussion rounds to finalise the code book. We also had fortnight meetings where the first author discussed the code book with other authors, allowing for peer review and feedback. 

\section{Conclusion} \label{SEC:Conclusion}
We conducted an empirical case study to understand how empathy is practised in the interactions between developers and end users. We employed an empathy test, a demographics questionnaire, an observation study and a set of in-depth interviews to collect data. Mixed methods were used including STGT for qualitative data and descriptive statistics for quantitative data analysis. We identified some enablers of empathy and the nature of awareness needed to trigger empathy. We determined some barriers to empathy and strategies that could be employed to overcome these barriers. Based on our findings, we report a set of actions that can be used to identify the types of awareness required to enable empathy, as well as a set of strategies to overcome empathy barriers. We identified verbal and nonverbal empathy cues demonstrated by participants during the observation sessions. We determined some trends using the scores of empathy test. We report insights on emerging relationships and differentiate empathy enablers based on cognitive and affective empathy. Extending the findings of this study will be beneficial for both software practitioners and research community. We presented a set of recommendations and potential future works for software practitioners and researchers.

\begin{acks}
Gunatilake, Grundy and Mueller are supported by ARC Laureate Fellowship FL190100035. We thank all our participants for their tremendous contribution and all the other stakeholders of the project including the MCHRI and MADA. 
\end{acks}

\bibliographystyle{ACM-Reference-Format}
\bibliography{manuscript}


\begin{thebibliography}{79}


\ifx \showCODEN    \undefined \def \showCODEN     #1{\unskip}     \fi
\ifx \showDOI      \undefined \def \showDOI       #1{#1}\fi
\ifx \showISBNx    \undefined \def \showISBNx     #1{\unskip}     \fi
\ifx \showISBNxiii \undefined \def \showISBNxiii  #1{\unskip}     \fi
\ifx \showISSN     \undefined \def \showISSN      #1{\unskip}     \fi
\ifx \showLCCN     \undefined \def \showLCCN      #1{\unskip}     \fi
\ifx \shownote     \undefined \def \shownote      #1{#1}          \fi
\ifx \showarticletitle \undefined \def \showarticletitle #1{#1}   \fi
\ifx \showURL      \undefined \def \showURL       {\relax}        \fi
\providecommand\bibfield[2]{#2}
\providecommand\bibinfo[2]{#2}
\providecommand\natexlab[1]{#1}
\providecommand\showeprint[2][]{arXiv:#2}

\bibitem[Airenti(2015)]%
        {airenti2015cognitive}
\bibfield{author}{\bibinfo{person}{Gabriella Airenti}.} \bibinfo{year}{2015}\natexlab{}.
\newblock \showarticletitle{The cognitive bases of anthropomorphism: from relatedness to empathy}.
\newblock \bibinfo{journal}{\emph{International Journal of Social Robotics}}  \bibinfo{volume}{7} (\bibinfo{year}{2015}), \bibinfo{pages}{117--127}.
\newblock


\bibitem[Akgün et~al\mbox{.}(2015)]%
        {akgun2015collectiveempathy}
\bibfield{author}{\bibinfo{person}{Ali~E. Akgün}, \bibinfo{person}{Halit Keskin}, \bibinfo{person}{A.~Yavuz Cebecioglu}, {and} \bibinfo{person}{Derya Dogan}.} \bibinfo{year}{2015}\natexlab{}.
\newblock \showarticletitle{Antecedents and consequences of collective empathy in software development project teams}.
\newblock \bibinfo{journal}{\emph{Information \& Management}} \bibinfo{volume}{52}, \bibinfo{number}{2} (\bibinfo{year}{2015}), \bibinfo{pages}{247--259}.
\newblock
\showISSN{0378-7206}
\urldef\tempurl%
\url{https://doi.org/10.1016/j.im.2014.11.004}
\showDOI{\tempurl}


\bibitem[Baron-Cohen and Wheelwright(2004)]%
        {baron2004EQ}
\bibfield{author}{\bibinfo{person}{Simon Baron-Cohen} {and} \bibinfo{person}{Sally Wheelwright}.} \bibinfo{year}{2004}\natexlab{}.
\newblock \showarticletitle{The empathy quotient: an investigation of adults with Asperger syndrome or high functioning autism, and normal sex differences}.
\newblock \bibinfo{journal}{\emph{Journal of autism and developmental disorders}}  \bibinfo{volume}{34} (\bibinfo{year}{2004}), \bibinfo{pages}{163--175}.
\newblock


\bibitem[Bartel and Saavedra(2000)]%
        {bartel2000collective}
\bibfield{author}{\bibinfo{person}{Caroline~A Bartel} {and} \bibinfo{person}{Richard Saavedra}.} \bibinfo{year}{2000}\natexlab{}.
\newblock \showarticletitle{The collective construction of work group moods}.
\newblock \bibinfo{journal}{\emph{Administrative Science Quarterly}} \bibinfo{volume}{45}, \bibinfo{number}{2} (\bibinfo{year}{2000}), \bibinfo{pages}{197--231}.
\newblock


\bibitem[Batson and Ahmad(2009)]%
        {batson2009intergroup}
\bibfield{author}{\bibinfo{person}{C~Daniel Batson} {and} \bibinfo{person}{Nadia~Y Ahmad}.} \bibinfo{year}{2009}\natexlab{}.
\newblock \showarticletitle{Using empathy to improve intergroup attitudes and relations}.
\newblock \bibinfo{journal}{\emph{Social issues and policy review}} \bibinfo{volume}{3}, \bibinfo{number}{1} (\bibinfo{year}{2009}), \bibinfo{pages}{141--177}.
\newblock


\bibitem[Batson et~al\mbox{.}(1996)]%
        {batson1996prior}
\bibfield{author}{\bibinfo{person}{C.~Daniel Batson}, \bibinfo{person}{Susie~C. Sympson}, \bibinfo{person}{Jennifer~L. Hindman}, \bibinfo{person}{Peter Decruz}, \bibinfo{person}{R.~Matthew Todd}, \bibinfo{person}{Joy~L. Weeks}, \bibinfo{person}{Geoffrey Jennings}, {and} \bibinfo{person}{Christopher~T. Burns}.} \bibinfo{year}{1996}\natexlab{}.
\newblock \showarticletitle{"I've Been there, Too": Effect on Empathy of Prior Experience with a Need}.
\newblock \bibinfo{journal}{\emph{Personality and Social Psychology Bulletin}} \bibinfo{volume}{22}, \bibinfo{number}{5} (\bibinfo{year}{1996}), \bibinfo{pages}{474--482}.
\newblock
\urldef\tempurl%
\url{https://doi.org/10.1177/0146167296225005}
\showDOI{\tempurl}
\showeprint{https://doi.org/10.1177/0146167296225005}


\bibitem[Blanco et~al\mbox{.}(2017)]%
        {blanco2017deconstructing}
\bibfield{author}{\bibinfo{person}{Teresa Blanco}, \bibinfo{person}{Ignacio L{\'o}pez-Forni{\'e}s}, {and} \bibinfo{person}{Francisco~Javier Zarazaga-Soria}.} \bibinfo{year}{2017}\natexlab{}.
\newblock \showarticletitle{Deconstructing the Tower of Babel: a design method to improve empathy and teamwork competences of informatics students}.
\newblock \bibinfo{journal}{\emph{International Journal of Technology and Design Education}} \bibinfo{volume}{27}, \bibinfo{number}{2} (\bibinfo{year}{2017}), \bibinfo{pages}{307--328}.
\newblock


\bibitem[Canedo et~al\mbox{.}(2020)]%
        {canedo2020design}
\bibfield{author}{\bibinfo{person}{Edna~Dias Canedo}, \bibinfo{person}{Ana Carolina Dos~Santos Pergentino}, \bibinfo{person}{Angelica Toffano~Seidel Calazans}, \bibinfo{person}{Frederico~Viana Almeida}, \bibinfo{person}{Pedro Henrique~Teixeira Costa}, {and} \bibinfo{person}{Fernanda Lima}.} \bibinfo{year}{2020}\natexlab{}.
\newblock \showarticletitle{Design Thinking Use in Agile Software Projects: Software Developers' Perception}. In \bibinfo{booktitle}{\emph{In Proceedings of the 22nd International Conference on Enterprise Information Systems (ICEIS2020)}}, Vol.~\bibinfo{volume}{978-989-758-423-7}. \bibinfo{publisher}{SCITEPRESS–Science and Technology Publications}, \bibinfo{pages}{217--224}.
\newblock
\showISBNx{978-989-758-423-7}
\urldef\tempurl%
\url{https://doi.org/10.5220/0009387502170224}
\showDOI{\tempurl}


\bibitem[Cassels et~al\mbox{.}(2010)]%
        {cassels2010role}
\bibfield{author}{\bibinfo{person}{Tracy~G Cassels}, \bibinfo{person}{Sherilynn Chan}, {and} \bibinfo{person}{Winnie Chung}.} \bibinfo{year}{2010}\natexlab{}.
\newblock \showarticletitle{The role of culture in affective empathy: Cultural and bicultural differences}.
\newblock \bibinfo{journal}{\emph{Journal of Cognition and Culture}} \bibinfo{volume}{10}, \bibinfo{number}{3-4} (\bibinfo{year}{2010}), \bibinfo{pages}{309--326}.
\newblock


\bibitem[Chartrand and Bargh(1999)]%
        {chartrand1999chameleon}
\bibfield{author}{\bibinfo{person}{Tanya~L Chartrand} {and} \bibinfo{person}{John~A Bargh}.} \bibinfo{year}{1999}\natexlab{}.
\newblock \showarticletitle{The chameleon effect: The perception--behavior link and social interaction.}
\newblock \bibinfo{journal}{\emph{Journal of personality and social psychology}} \bibinfo{volume}{76}, \bibinfo{number}{6} (\bibinfo{year}{1999}), \bibinfo{pages}{893}.
\newblock


\bibitem[Chopik et~al\mbox{.}(2017)]%
        {chopik2017differences}
\bibfield{author}{\bibinfo{person}{William~J Chopik}, \bibinfo{person}{Ed O’Brien}, {and} \bibinfo{person}{Sara~H Konrath}.} \bibinfo{year}{2017}\natexlab{}.
\newblock \showarticletitle{Differences in empathic concern and perspective taking across 63 countries}.
\newblock \bibinfo{journal}{\emph{Journal of Cross-Cultural Psychology}} \bibinfo{volume}{48}, \bibinfo{number}{1} (\bibinfo{year}{2017}), \bibinfo{pages}{23--38}.
\newblock


\bibitem[Clark et~al\mbox{.}(2019)]%
        {clark2019feel}
\bibfield{author}{\bibinfo{person}{Malissa~A Clark}, \bibinfo{person}{Melissa~M Robertson}, {and} \bibinfo{person}{Stephen Young}.} \bibinfo{year}{2019}\natexlab{}.
\newblock \showarticletitle{\"I feel your pain": A critical review of organizational research on empathy}.
\newblock \bibinfo{journal}{\emph{Journal of Organizational Behavior}} \bibinfo{volume}{40}, \bibinfo{number}{2} (\bibinfo{year}{2019}), \bibinfo{pages}{166--192}.
\newblock


\bibitem[Cuff et~al\mbox{.}(2016)]%
        {cuff2016empathy}
\bibfield{author}{\bibinfo{person}{Benjamin~MP Cuff}, \bibinfo{person}{Sarah~J Brown}, \bibinfo{person}{Laura Taylor}, {and} \bibinfo{person}{Douglas~J Howat}.} \bibinfo{year}{2016}\natexlab{}.
\newblock \showarticletitle{Empathy: A review of the concept}.
\newblock \bibinfo{journal}{\emph{Emotion review}} \bibinfo{volume}{8}, \bibinfo{number}{2} (\bibinfo{year}{2016}), \bibinfo{pages}{144--153}.
\newblock


\bibitem[Davis(1980)]%
        {davis1980IRI}
\bibfield{author}{\bibinfo{person}{Mark Davis}.} \bibinfo{year}{1980}\natexlab{}.
\newblock \showarticletitle{A multidimensional approach to individual differences in empathy}.
\newblock \bibinfo{journal}{\emph{JSAS Catalog of Selected Documents in Psychology}}  \bibinfo{volume}{10} (\bibinfo{year}{1980}).
\newblock


\bibitem[Davis(1983)]%
        {davis1983IRI}
\bibfield{author}{\bibinfo{person}{Mark~H Davis}.} \bibinfo{year}{1983}\natexlab{}.
\newblock \showarticletitle{Measuring individual differences in empathy: Evidence for a multidimensional approach.}
\newblock \bibinfo{journal}{\emph{Journal of personality and social psychology}} \bibinfo{volume}{44}, \bibinfo{number}{1} (\bibinfo{year}{1983}), \bibinfo{pages}{113}.
\newblock


\bibitem[De~Jong and Lentz(2007)]%
        {de2007professional}
\bibfield{author}{\bibinfo{person}{Menno De~Jong} {and} \bibinfo{person}{Leo Lentz}.} \bibinfo{year}{2007}\natexlab{}.
\newblock \showarticletitle{Professional writers and empathy: Exploring the barriers to anticipating reader problems}. In \bibinfo{booktitle}{\emph{2007 IEEE International Professional Communication Conference}}. \bibinfo{publisher}{IEEE}, \bibinfo{address}{Seattle, WA, USA}, \bibinfo{pages}{1--8}.
\newblock
\urldef\tempurl%
\url{https://doi.org/10.1109/IPCC.2007.4464058}
\showDOI{\tempurl}


\bibitem[De~Waal(2008)]%
        {de2008putting}
\bibfield{author}{\bibinfo{person}{Frans~BM De~Waal}.} \bibinfo{year}{2008}\natexlab{}.
\newblock \showarticletitle{Putting the altruism back into altruism: The evolution of empathy}.
\newblock \bibinfo{journal}{\emph{Annu. Rev. Psychol.}}  \bibinfo{volume}{59} (\bibinfo{year}{2008}), \bibinfo{pages}{279--300}.
\newblock


\bibitem[Decety and Jackson(2004)]%
        {decety2004functional}
\bibfield{author}{\bibinfo{person}{Jean Decety} {and} \bibinfo{person}{Philip~L. Jackson}.} \bibinfo{year}{2004}\natexlab{}.
\newblock \showarticletitle{The Functional Architecture of Human Empathy}.
\newblock \bibinfo{journal}{\emph{Behavioral and Cognitive Neuroscience Reviews}} \bibinfo{volume}{3}, \bibinfo{number}{2} (\bibinfo{year}{2004}), \bibinfo{pages}{71--100}.
\newblock
\urldef\tempurl%
\url{https://doi.org/10.1177/1534582304267187}
\showDOI{\tempurl}
\showeprint{https://doi.org/10.1177/1534582304267187}
\newblock
\shownote{PMID: 15537986}.


\bibitem[Decety and Moriguchi(2007)]%
        {decety2007empathic}
\bibfield{author}{\bibinfo{person}{Jean Decety} {and} \bibinfo{person}{Yoshiya Moriguchi}.} \bibinfo{year}{2007}\natexlab{}.
\newblock \showarticletitle{The empathic brain and its dysfunction in psychiatric populations: Implications for intervention across different clinical conditions}.
\newblock \bibinfo{journal}{\emph{BioPsychoSocial medicine}} \bibinfo{volume}{1}, \bibinfo{number}{1} (\bibinfo{year}{2007}), \bibinfo{pages}{1--21}.
\newblock


\bibitem[Delpechitre(2013)]%
        {delpechitre2013review}
\bibfield{author}{\bibinfo{person}{Duleep Delpechitre}.} \bibinfo{year}{2013}\natexlab{}.
\newblock \showarticletitle{Review and assessment of past empathy scales to measure salesperson’s empathy}.
\newblock \bibinfo{journal}{\emph{Journal of Management and Marketing Research}} \bibinfo{volume}{13}, \bibinfo{number}{1} (\bibinfo{year}{2013}), \bibinfo{pages}{1--16}.
\newblock


\bibitem[Derksen et~al\mbox{.}(2016)]%
        {derksen2016managing}
\bibfield{author}{\bibinfo{person}{Frans~AWM Derksen}, \bibinfo{person}{Tim~C olde Hartman}, \bibinfo{person}{Jozien~M Bensing}, {and} \bibinfo{person}{Antoine~LM Lagro-Janssen}.} \bibinfo{year}{2016}\natexlab{}.
\newblock \showarticletitle{Managing barriers to empathy in the clinical encounter: a qualitative interview study with GPs}.
\newblock \bibinfo{journal}{\emph{British Journal of General Practice}} \bibinfo{volume}{66}, \bibinfo{number}{653} (\bibinfo{year}{2016}), \bibinfo{pages}{e887--e895}.
\newblock


\bibitem[Eisenberg and Eggum(2009)]%
        {eisenberg2009empathic}
\bibfield{author}{\bibinfo{person}{Nancy Eisenberg} {and} \bibinfo{person}{Natalie~D Eggum}.} \bibinfo{year}{2009}\natexlab{}.
\newblock \showarticletitle{Empathic responding: Sympathy and personal distress}.
\newblock \bibinfo{journal}{\emph{The social neuroscience of empathy}} \bibinfo{volume}{6}, \bibinfo{number}{2009} (\bibinfo{year}{2009}), \bibinfo{pages}{71--830}.
\newblock


\bibitem[Eisenberg et~al\mbox{.}(2014)]%
        {eisenberg2014sympathy}
\bibfield{author}{\bibinfo{person}{Nancy Eisenberg}, \bibinfo{person}{Tracy~L Spinrad}, {and} \bibinfo{person}{Zoe~E Taylor}.} \bibinfo{year}{2014}\natexlab{}.
\newblock \showarticletitle{Sympathy}.
\newblock In \bibinfo{booktitle}{\emph{The Handbook of Virtue Ethics} (\bibinfo{edition}{1} ed.)}. \bibinfo{publisher}{Routledge}, \bibinfo{address}{London}, \bibinfo{pages}{409--417}.
\newblock
\urldef\tempurl%
\url{https://doi.org/10.4324/9781315729053}
\showDOI{\tempurl}


\bibitem[EKLUND et~al\mbox{.}(2009)]%
        {eklund2009similar}
\bibfield{author}{\bibinfo{person}{JAKOB EKLUND}, \bibinfo{person}{TERESIA ANDERSSON-STRÅBERG}, {and} \bibinfo{person}{ERIC~M. HANSEN}.} \bibinfo{year}{2009}\natexlab{}.
\newblock \showarticletitle{“I've also experienced loss and fear”: Effects of prior similar experience on empathy}.
\newblock \bibinfo{journal}{\emph{Scandinavian Journal of Psychology}} \bibinfo{volume}{50}, \bibinfo{number}{1} (\bibinfo{year}{2009}), \bibinfo{pages}{65--69}.
\newblock
\urldef\tempurl%
\url{https://doi.org/10.1111/j.1467-9450.2008.00673.x}
\showDOI{\tempurl}
\showeprint{https://onlinelibrary.wiley.com/doi/pdf/10.1111/j.1467-9450.2008.00673.x}


\bibitem[Ekman and Friesen(1971)]%
        {ekman1971constants}
\bibfield{author}{\bibinfo{person}{Paul Ekman} {and} \bibinfo{person}{Wallace~V Friesen}.} \bibinfo{year}{1971}\natexlab{}.
\newblock \showarticletitle{Constants across cultures in the face and emotion.}
\newblock \bibinfo{journal}{\emph{Journal of personality and social psychology}} \bibinfo{volume}{17}, \bibinfo{number}{2} (\bibinfo{year}{1971}), \bibinfo{pages}{124}.
\newblock


\bibitem[Elliott et~al\mbox{.}(2011)]%
        {elliott2011empathy}
\bibfield{author}{\bibinfo{person}{Robert Elliott}, \bibinfo{person}{Arthur~C Bohart}, \bibinfo{person}{Jeanne~C Watson}, {and} \bibinfo{person}{Leslie~S Greenberg}.} \bibinfo{year}{2011}\natexlab{}.
\newblock \showarticletitle{Empathy.}
\newblock \bibinfo{journal}{\emph{Psychotherapy}} \bibinfo{volume}{48}, \bibinfo{number}{1} (\bibinfo{year}{2011}), \bibinfo{pages}{43}.
\newblock


\bibitem[Ewin et~al\mbox{.}(2021)]%
        {ewin2021empathy}
\bibfield{author}{\bibinfo{person}{Natalie Ewin}, \bibinfo{person}{Ritesh Chugh}, \bibinfo{person}{Olav Muurlink}, \bibinfo{person}{Jacqueline Jarvis}, {and} \bibinfo{person}{Jo Luck}.} \bibinfo{year}{2021}\natexlab{}.
\newblock \showarticletitle{Empathy of project management students and why it matters}.
\newblock \bibinfo{journal}{\emph{Procedia Computer Science}}  \bibinfo{volume}{181} (\bibinfo{year}{2021}), \bibinfo{pages}{503--510}.
\newblock


\bibitem[Ferreira et~al\mbox{.}(2015)]%
        {ferreira2015designing}
\bibfield{author}{\bibinfo{person}{Bruna Ferreira}, \bibinfo{person}{Williamson Silva}, \bibinfo{person}{Edson Oliveira}, {and} \bibinfo{person}{Tayana Conte}.} \bibinfo{year}{2015}\natexlab{}.
\newblock \showarticletitle{Designing Personas with Empathy Map}. In \bibinfo{booktitle}{\emph{Proceedings of the International Conference on Software Engineering and Knowledge Engineering, SEKE}}, Vol.~\bibinfo{volume}{152}. \bibinfo{address}{Pittsburgh, USA}.
\newblock
\showISSN{23259086}
\urldef\tempurl%
\url{https://doi.org/10.18293/SEKE2015-152}
\showDOI{\tempurl}


\bibitem[Ferreira et~al\mbox{.}(2016)]%
        {ferreira2016pathy}
\bibfield{author}{\bibinfo{person}{Bruna~Moraes Ferreira}, \bibinfo{person}{Simone~DJ Barbosa}, {and} \bibinfo{person}{Tayana Conte}.} \bibinfo{year}{2016}\natexlab{}.
\newblock \showarticletitle{PATHY: Using empathy with personas to design applications that meet the users’ needs}. In \bibinfo{booktitle}{\emph{International Conference on Human-Computer Interaction}}. \bibinfo{publisher}{Springer}, \bibinfo{address}{Toronto, ON, Canada}, \bibinfo{pages}{153--165}.
\newblock


\bibitem[Goldman(2011)]%
        {goldman2011two}
\bibfield{author}{\bibinfo{person}{Alvin~I. Goldman}.} \bibinfo{year}{2011}\natexlab{}.
\newblock \showarticletitle{{313 Two Routes to Empathy: Insights from Cognitive Neuroscience}}.
\newblock In \bibinfo{booktitle}{\emph{{Empathy: Philosophical and Psychological Perspectives}}}. \bibinfo{publisher}{Oxford University Press}, \bibinfo{address}{Oxford}, \bibinfo{pages}{31--44}.
\newblock
\showISBNx{9780199539956}
\urldef\tempurl%
\url{https://doi.org/10.1093/acprof:oso/9780199539956.003.0004}
\showDOI{\tempurl}
\showeprint{https://academic.oup.com/book/0/chapter/151861762/chapter-ag-pdf/44974436/book\_7216\_section\_151861762.ag.pdf}


\bibitem[Goleman(1996)]%
        {goleman1996emotionalintelligence}
\bibfield{author}{\bibinfo{person}{Daniel Goleman}.} \bibinfo{year}{1996}\natexlab{}.
\newblock \bibinfo{booktitle}{\emph{Emotional Intelligence why it can matter more IQ}}. Vol.~\bibinfo{volume}{53}.
\newblock \bibinfo{publisher}{Bloomsbury Publishing}, \bibinfo{address}{Great Britain}. 1689--1699 pages.
\newblock
Issue 9.
\showISBNx{9788578110796}
\showISSN{1098-6596}


\bibitem[Goleman et~al\mbox{.}(2002)]%
        {goleman2002emotional}
\bibfield{author}{\bibinfo{person}{Daniel Goleman}, \bibinfo{person}{Richard Boyatzis}, {and} \bibinfo{person}{Annie McKee}.} \bibinfo{year}{2002}\natexlab{}.
\newblock \showarticletitle{The emotional reality of teams}.
\newblock \bibinfo{journal}{\emph{Journal of Organizational Excellence}} \bibinfo{volume}{21}, \bibinfo{number}{2} (\bibinfo{year}{2002}), \bibinfo{pages}{55--65}.
\newblock


\bibitem[Graetsch et~al\mbox{.}(2023)]%
        {graetsch2023data}
\bibfield{author}{\bibinfo{person}{Ulrike~M. Graetsch}, \bibinfo{person}{Hourieh Khalajzadeh}, \bibinfo{person}{Mojtaba Shahin}, \bibinfo{person}{Rashina Hoda}, {and} \bibinfo{person}{John Grundy}.} \bibinfo{year}{2023}\natexlab{}.
\newblock \showarticletitle{Dealing With Data Challenges When Delivering Data-Intensive Software Solutions}.
\newblock \bibinfo{journal}{\emph{IEEE Transactions on Software Engineering}} (\bibinfo{year}{2023}), \bibinfo{pages}{1--23}.
\newblock
\urldef\tempurl%
\url{https://doi.org/10.1109/TSE.2023.3291003}
\showDOI{\tempurl}


\bibitem[Group et~al\mbox{.}(1999)]%
        {medical1999learning}
\bibfield{author}{\bibinfo{person}{Medical School Objectives~Writing Group} {et~al\mbox{.}}} \bibinfo{year}{1999}\natexlab{}.
\newblock \showarticletitle{Learning objectives for medical student education—guidelines for medical schools: report I of the Medical School Objectives Project}.
\newblock \bibinfo{journal}{\emph{Academic Medicine}} \bibinfo{volume}{74}, \bibinfo{number}{1} (\bibinfo{year}{1999}), \bibinfo{pages}{13--18}.
\newblock


\bibitem[Gunatilake et~al\mbox{.}(2023)]%
        {gunatilake2023empathy}
\bibfield{author}{\bibinfo{person}{Hashini Gunatilake}, \bibinfo{person}{John Grundy}, \bibinfo{person}{Ingo Mueller}, {and} \bibinfo{person}{Rashina Hoda}.} \bibinfo{year}{2023}\natexlab{}.
\newblock \showarticletitle{Empathy models and software engineering—A preliminary analysis and taxonomy}.
\newblock \bibinfo{journal}{\emph{Journal of Systems and Software}}  \bibinfo{volume}{203} (\bibinfo{year}{2023}), \bibinfo{pages}{111747}.
\newblock


\bibitem[Guthridge and Giummarra(2021)]%
        {guthridge2021taxonomy}
\bibfield{author}{\bibinfo{person}{Michaela Guthridge} {and} \bibinfo{person}{Melita~J Giummarra}.} \bibinfo{year}{2021}\natexlab{}.
\newblock \showarticletitle{The taxonomy of empathy: A meta-definition and the nine dimensions of the empathic system}.
\newblock \bibinfo{journal}{\emph{Journal of Humanistic Psychology}}  \bibinfo{volume}{0} (\bibinfo{year}{2021}), \bibinfo{pages}{00221678211018015}.
\newblock
\urldef\tempurl%
\url{https://doi.org/10.1177/00221678211018015}
\showDOI{\tempurl}


\bibitem[Halabi et~al\mbox{.}(2008)]%
        {halabi2008social}
\bibfield{author}{\bibinfo{person}{Samer Halabi}, \bibinfo{person}{John~F Dovidio}, {and} \bibinfo{person}{Arie Nadler}.} \bibinfo{year}{2008}\natexlab{}.
\newblock \showarticletitle{When and how do high status group members offer help: Effects of social dominance orientation and status threat}.
\newblock \bibinfo{journal}{\emph{Political Psychology}} \bibinfo{volume}{29}, \bibinfo{number}{6} (\bibinfo{year}{2008}), \bibinfo{pages}{841--858}.
\newblock


\bibitem[Halpern(2003)]%
        {halpern2003clinical}
\bibfield{author}{\bibinfo{person}{Jodi Halpern}.} \bibinfo{year}{2003}\natexlab{}.
\newblock \showarticletitle{What is clinical empathy?}
\newblock \bibinfo{journal}{\emph{Journal of general internal medicine}}  \bibinfo{volume}{18} (\bibinfo{year}{2003}), \bibinfo{pages}{670--674}.
\newblock


\bibitem[Hazel et~al\mbox{.}(2011)]%
        {hazel2011animalscience}
\bibfield{author}{\bibinfo{person}{Susan~J Hazel}, \bibinfo{person}{Tania~D Signal}, {and} \bibinfo{person}{Nicola Taylor}.} \bibinfo{year}{2011}\natexlab{}.
\newblock \showarticletitle{Can teaching veterinary and animal-science students about animal welfare affect their attitude toward animals and human-related empathy?}
\newblock \bibinfo{journal}{\emph{Journal of veterinary medical education}} \bibinfo{volume}{38}, \bibinfo{number}{1} (\bibinfo{year}{2011}), \bibinfo{pages}{74--83}.
\newblock


\bibitem[Heinke and Louis(2009)]%
        {heinke2009cultural}
\bibfield{author}{\bibinfo{person}{Miriam~S. Heinke} {and} \bibinfo{person}{Winnifred~R. Louis}.} \bibinfo{year}{2009}\natexlab{}.
\newblock \showarticletitle{Cultural Background and Individualistic–Collectivistic Values in Relation to Similarity, Perspective Taking, and Empathy}.
\newblock \bibinfo{journal}{\emph{Journal of Applied Social Psychology}} \bibinfo{volume}{39}, \bibinfo{number}{11} (\bibinfo{year}{2009}), \bibinfo{pages}{2570--2590}.
\newblock
\urldef\tempurl%
\url{https://doi.org/10.1111/j.1559-1816.2009.00538.x}
\showDOI{\tempurl}
\showeprint{https://onlinelibrary.wiley.com/doi/pdf/10.1111/j.1559-1816.2009.00538.x}


\bibitem[Hoda(2021)]%
        {hoda2021STGT}
\bibfield{author}{\bibinfo{person}{Rashina Hoda}.} \bibinfo{year}{2021}\natexlab{}.
\newblock \showarticletitle{Socio-technical grounded theory for software engineering}.
\newblock \bibinfo{journal}{\emph{IEEE Transactions on Software Engineering}} \bibinfo{volume}{48}, \bibinfo{number}{10} (\bibinfo{year}{2021}), \bibinfo{pages}{3808--3832}.
\newblock


\bibitem[Hoda et~al\mbox{.}(2011)]%
        {hoda2011selforganizingagile}
\bibfield{author}{\bibinfo{person}{Rashina Hoda}, \bibinfo{person}{James Noble}, {and} \bibinfo{person}{Stuart Marshall}.} \bibinfo{year}{2011}\natexlab{}.
\newblock \showarticletitle{The impact of inadequate customer collaboration on self-organizing Agile teams}.
\newblock \bibinfo{journal}{\emph{Information and software technology}} \bibinfo{volume}{53}, \bibinfo{number}{5} (\bibinfo{year}{2011}), \bibinfo{pages}{521--534}.
\newblock


\bibitem[Hoffman(2001)]%
        {hoffman2001empathy}
\bibfield{author}{\bibinfo{person}{Martin~L Hoffman}.} \bibinfo{year}{2001}\natexlab{}.
\newblock \bibinfo{booktitle}{\emph{Empathy and moral development: Implications for caring and justice}}.
\newblock \bibinfo{publisher}{Cambridge University Press}, \bibinfo{address}{Cambridge, UK}.
\newblock


\bibitem[Hofstede(1991)]%
        {hofstede1991empirical}
\bibfield{author}{\bibinfo{person}{Geert Hofstede}.} \bibinfo{year}{1991}\natexlab{}.
\newblock \bibinfo{booktitle}{\emph{Empirical models of cultural differences}}.
\newblock \bibinfo{publisher}{Swets \& Zeitlinger Publishers}, \bibinfo{address}{Lisse, Netherlands}, \bibinfo{pages}{4--20}.
\newblock
\showISBNx{90-265-1117-5}


\bibitem[Hojat(2016)]%
        {hojat2016empathy}
\bibfield{author}{\bibinfo{person}{Mohammadreza Hojat}.} \bibinfo{year}{2016}\natexlab{}.
\newblock \bibinfo{booktitle}{\emph{Empathy in health professions education and patient care}}.
\newblock \bibinfo{publisher}{Springer}, \bibinfo{address}{Switzerland}.
\newblock
\urldef\tempurl%
\url{https://doi.org/10.1007/978-3-319-27625-0}
\showDOI{\tempurl}


\bibitem[Hojat et~al\mbox{.}(2018)]%
        {hojat2018jefferson}
\bibfield{author}{\bibinfo{person}{Mohammadreza Hojat}, \bibinfo{person}{Jennifer DeSantis}, \bibinfo{person}{Stephen~C Shannon}, \bibinfo{person}{Luke~H Mortensen}, \bibinfo{person}{Mark~R Speicher}, \bibinfo{person}{Lynn Bragan}, \bibinfo{person}{Marianna LaNoue}, {and} \bibinfo{person}{Leonard~H Calabrese}.} \bibinfo{year}{2018}\natexlab{}.
\newblock \showarticletitle{The Jefferson Scale of Empathy: a nationwide study of measurement properties, underlying components, latent variable structure, and national norms in medical students}.
\newblock \bibinfo{journal}{\emph{Advances in Health Sciences Education}} \bibinfo{volume}{23}, \bibinfo{number}{5} (\bibinfo{year}{2018}), \bibinfo{pages}{899--920}.
\newblock


\bibitem[Hojat et~al\mbox{.}(2001)]%
        {hojat2001JSE}
\bibfield{author}{\bibinfo{person}{Mohammadreza Hojat}, \bibinfo{person}{Salvatore Mangione}, \bibinfo{person}{Thomas~J Nasca}, \bibinfo{person}{Mitchell~JM Cohen}, \bibinfo{person}{Joseph~S Gonnella}, \bibinfo{person}{James~B Erdmann}, \bibinfo{person}{Jon Veloski}, {and} \bibinfo{person}{Mike Magee}.} \bibinfo{year}{2001}\natexlab{}.
\newblock \showarticletitle{The Jefferson Scale of Physician Empathy: development and preliminary psychometric data}.
\newblock \bibinfo{journal}{\emph{Educational and psychological measurement}} \bibinfo{volume}{61}, \bibinfo{number}{2} (\bibinfo{year}{2001}), \bibinfo{pages}{349--365}.
\newblock


\bibitem[Howick and Rees(2017)]%
        {howick2017barriers}
\bibfield{author}{\bibinfo{person}{J Howick} {and} \bibinfo{person}{S Rees}.} \bibinfo{year}{2017}\natexlab{}.
\newblock \showarticletitle{Overthrowing barriers to empathy in healthcare: empathy in the age of the Internet}.
\newblock \bibinfo{journal}{\emph{Journal of the Royal Society of Medicine}} \bibinfo{volume}{110}, \bibinfo{number}{9} (\bibinfo{year}{2017}), \bibinfo{pages}{352--357}.
\newblock
\urldef\tempurl%
\url{https://doi.org/10.1177/0141076817714443}
\showDOI{\tempurl}
\showeprint{https://doi.org/10.1177/0141076817714443}


\bibitem[Ilgunaite et~al\mbox{.}(2017)]%
        {ilgunaite2017measuring}
\bibfield{author}{\bibinfo{person}{Guste Ilgunaite}, \bibinfo{person}{Luciano Giromini}, {and} \bibinfo{person}{Marzia Di~Girolamo}.} \bibinfo{year}{2017}\natexlab{}.
\newblock \showarticletitle{Measuring empathy: A literature review of available tools.}
\newblock \bibinfo{journal}{\emph{BPA-Applied Psychology Bulletin (Bollettino di Psicologia Applicata)}} \bibinfo{volume}{65}, \bibinfo{number}{280} (\bibinfo{year}{2017}), \bibinfo{pages}{2--28}.
\newblock


\bibitem[Jami et~al\mbox{.}(2023)]%
        {jami2023interaction}
\bibfield{author}{\bibinfo{person}{Parvaneh~Yaghoubi Jami}, \bibinfo{person}{David~Ian Walker}, {and} \bibinfo{person}{Behzad Mansouri}.} \bibinfo{year}{2023}\natexlab{}.
\newblock \showarticletitle{Interaction of empathy and culture: a review}.
\newblock \bibinfo{journal}{\emph{Current Psychology}}  \bibinfo{volume}{42} (\bibinfo{year}{2023}), \bibinfo{pages}{1--16}.
\newblock
\urldef\tempurl%
\url{https://doi.org/10.1007/s12144-023-04422-6}
\showDOI{\tempurl}


\bibitem[Karolita et~al\mbox{.}(2023)]%
        {karolita2023personas}
\bibfield{author}{\bibinfo{person}{Devi Karolita}, \bibinfo{person}{Jennifer McIntosh}, \bibinfo{person}{Tanjila Kanij}, \bibinfo{person}{John Grundy}, {and} \bibinfo{person}{Humphrey~O. Obie}.} \bibinfo{year}{2023}\natexlab{}.
\newblock \showarticletitle{Use of personas in Requirements Engineering: A systematic mapping study}.
\newblock \bibinfo{journal}{\emph{Information and Software Technology}}  \bibinfo{volume}{162} (\bibinfo{year}{2023}), \bibinfo{pages}{107--264}.
\newblock
\showISSN{0950-5849}
\urldef\tempurl%
\url{https://doi.org/10.1016/j.infsof.2023.107264}
\showDOI{\tempurl}


\bibitem[Kitayama et~al\mbox{.}(2000)]%
        {kitayama2000culture}
\bibfield{author}{\bibinfo{person}{Shinobu Kitayama}, \bibinfo{person}{Hazel~Rose Markus}, {and} \bibinfo{person}{Masaru Kurokawa}.} \bibinfo{year}{2000}\natexlab{}.
\newblock \showarticletitle{Culture, emotion, and well-being: Good feelings in Japan and the United States}.
\newblock \bibinfo{journal}{\emph{Cognition \& Emotion}} \bibinfo{volume}{14}, \bibinfo{number}{1} (\bibinfo{year}{2000}), \bibinfo{pages}{93--124}.
\newblock


\bibitem[Levy(2018)]%
        {levy2018educating}
\bibfield{author}{\bibinfo{person}{Meira Levy}.} \bibinfo{year}{2018}\natexlab{}.
\newblock \showarticletitle{Educating for Empathy in Software Engineering Course}. In \bibinfo{booktitle}{\emph{Joint Proceedings of {REFSQ-2018} Workshops, Doctoral Symposium, Live Studies Track, and Poster Track co-located with the 23rd International Conference on Requirements Engineering: Foundation for Software Quality {(REFSQ} 2018), Utrecht, The Netherlands, March 19, 2018}} \emph{(\bibinfo{series}{{CEUR} Workshop Proceedings}, Vol.~\bibinfo{volume}{2075})}, \bibfield{editor}{\bibinfo{person}{Klaus Schmid}, \bibinfo{person}{Paola Spoletini}, \bibinfo{person}{Eya~Ben Charrada}, \bibinfo{person}{Yoram Chisik}, \bibinfo{person}{Fabiano Dalpiaz}, \bibinfo{person}{Alessio Ferrari}, \bibinfo{person}{Peter Forbrig}, \bibinfo{person}{Xavier Franch}, \bibinfo{person}{Marite Kirikova}, \bibinfo{person}{Nazim~H. Madhavji}, \bibinfo{person}{Cristina Palomares}, \bibinfo{person}{Jolita Ralyt{\'{e}}}, \bibinfo{person}{Mehrdad Sabetzadeh}, \bibinfo{person}{Pete Sawyer}, \bibinfo{person}{Dirk van~der Linden}, {and}
  \bibinfo{person}{Anna Zamansky}} (Eds.). \bibinfo{publisher}{CEUR-WS.org}, \bibinfo{address}{Utrecht, The Netherlands}, \bibinfo{pages}{1--9}.
\newblock
\urldef\tempurl%
\url{https://ceur-ws.org/Vol-2075/FIRE18\_paper2.pdf}
\showURL{%
\tempurl}


\bibitem[Levy and Hadar(2018)]%
        {levy2018importance}
\bibfield{author}{\bibinfo{person}{Meira Levy} {and} \bibinfo{person}{Irit Hadar}.} \bibinfo{year}{2018}\natexlab{}.
\newblock \showarticletitle{The importance of empathy for analyzing privacy requirements}. In \bibinfo{booktitle}{\emph{2018 IEEE 5th International Workshop on Evolving Security \& Privacy Requirements Engineering (ESPRE)}}. \bibinfo{publisher}{IEEE}, \bibinfo{address}{Banff, Canada}, \bibinfo{pages}{9--13}.
\newblock
\urldef\tempurl%
\url{https://doi.org/10.1109/ESPRE.2018.00008}
\showDOI{\tempurl}


\bibitem[McAvoy and Butler(2009)]%
        {mcavoy2009failure}
\bibfield{author}{\bibinfo{person}{John McAvoy} {and} \bibinfo{person}{Tom Butler}.} \bibinfo{year}{2009}\natexlab{}.
\newblock \showarticletitle{A failure to learn by software developers: Inhibiting the adoption of an agile software development methodology}.
\newblock \bibinfo{journal}{\emph{Journal of Information Technology Case and Application Research}} \bibinfo{volume}{11}, \bibinfo{number}{1} (\bibinfo{year}{2009}), \bibinfo{pages}{23--46}.
\newblock


\bibitem[Motomura et~al\mbox{.}(2015)]%
        {motomura2015interaction}
\bibfield{author}{\bibinfo{person}{Yuki Motomura}, \bibinfo{person}{Akira Takeshita}, \bibinfo{person}{Yuka Egashira}, \bibinfo{person}{Takayuki Nishimura}, \bibinfo{person}{Yeon-kyu Kim}, {and} \bibinfo{person}{Shigeki Watanuki}.} \bibinfo{year}{2015}\natexlab{}.
\newblock \showarticletitle{Interaction between valence of empathy and familiarity: is it difficult to empathize with the positive events of a stranger?}
\newblock \bibinfo{journal}{\emph{Journal of physiological anthropology}}  \bibinfo{volume}{34} (\bibinfo{year}{2015}), \bibinfo{pages}{1--9}.
\newblock


\bibitem[Neumann et~al\mbox{.}(2015)]%
        {neumann2015measures}
\bibfield{author}{\bibinfo{person}{David.~L. Neumann}, \bibinfo{person}{Raymond~C.K. Chan}, \bibinfo{person}{Gregory.~J. Boyle}, \bibinfo{person}{Yi Wang}, {and} \bibinfo{person}{H. {Rae Westbury}}.} \bibinfo{year}{2015}\natexlab{}.
\newblock \showarticletitle{Chapter 10 - Measures of Empathy: Self-Report, Behavioral, and Neuroscientific Approaches}.
\newblock In \bibinfo{booktitle}{\emph{Measures of Personality and Social Psychological Constructs}}, \bibfield{editor}{\bibinfo{person}{Gregory~J. Boyle}, \bibinfo{person}{Donald~H. Saklofske}, {and} \bibinfo{person}{Gerald Matthews}} (Eds.). \bibinfo{publisher}{Academic Press}, \bibinfo{address}{San Diego}, \bibinfo{pages}{257--289}.
\newblock
\showISBNx{978-0-12-386915-9}
\urldef\tempurl%
\url{https://doi.org/10.1016/B978-0-12-386915-9.00010-3}
\showDOI{\tempurl}


\bibitem[Nicolai et~al\mbox{.}(2007)]%
        {nicolai2007rating}
\bibfield{author}{\bibinfo{person}{Jennifer Nicolai}, \bibinfo{person}{Ralf Demmel}, {and} \bibinfo{person}{Jutta Hagen}.} \bibinfo{year}{2007}\natexlab{}.
\newblock \showarticletitle{Rating scales for the assessment of empathic communication in medical interviews (REM): Scale development, reliability, and validity}.
\newblock \bibinfo{journal}{\emph{Journal of Clinical Psychology in Medical Settings}}  \bibinfo{volume}{14} (\bibinfo{year}{2007}), \bibinfo{pages}{367--375}.
\newblock


\bibitem[Nunes et~al\mbox{.}(2011)]%
        {nunes2011healthdisciplines}
\bibfield{author}{\bibinfo{person}{Paula Nunes}, \bibinfo{person}{Stella Williams}, \bibinfo{person}{Bidyadhar Sa}, {and} \bibinfo{person}{Keith Stevenson}.} \bibinfo{year}{2011}\natexlab{}.
\newblock \showarticletitle{A study of empathy decline in students from five health disciplines during their first year of training}.
\newblock \bibinfo{journal}{\emph{Int J Med Educ}}  \bibinfo{volume}{2} (\bibinfo{year}{2011}), \bibinfo{pages}{12--17}.
\newblock


\bibitem[Preis and Kroener-Herwig(2012)]%
        {preis2012pain}
\bibfield{author}{\bibinfo{person}{M.A. Preis} {and} \bibinfo{person}{B. Kroener-Herwig}.} \bibinfo{year}{2012}\natexlab{}.
\newblock \showarticletitle{Empathy for pain: The effects of prior experience and sex}.
\newblock \bibinfo{journal}{\emph{European Journal of Pain}} \bibinfo{volume}{16}, \bibinfo{number}{9} (\bibinfo{year}{2012}), \bibinfo{pages}{1311--1319}.
\newblock
\urldef\tempurl%
\url{https://doi.org/10.1002/j.1532-2149.2012.00119.x}
\showDOI{\tempurl}
\showeprint{https://onlinelibrary.wiley.com/doi/pdf/10.1002/j.1532-2149.2012.00119.x}


\bibitem[Preston and De~Waal(2002)]%
        {preston2002empathy}
\bibfield{author}{\bibinfo{person}{Stephanie~D Preston} {and} \bibinfo{person}{Frans~BM De~Waal}.} \bibinfo{year}{2002}\natexlab{}.
\newblock \showarticletitle{Empathy: Its ultimate and proximate bases}.
\newblock \bibinfo{journal}{\emph{Behavioral and brain sciences}} \bibinfo{volume}{25}, \bibinfo{number}{1} (\bibinfo{year}{2002}), \bibinfo{pages}{1--20}.
\newblock


\bibitem[Reniers et~al\mbox{.}(2011)]%
        {reniers2011QCAE}
\bibfield{author}{\bibinfo{person}{Renate~LEP Reniers}, \bibinfo{person}{Rhiannon Corcoran}, \bibinfo{person}{Richard Drake}, \bibinfo{person}{Nick~M Shryane}, {and} \bibinfo{person}{Birgit~A V{\"o}llm}.} \bibinfo{year}{2011}\natexlab{}.
\newblock \showarticletitle{The QCAE: A questionnaire of cognitive and affective empathy}.
\newblock \bibinfo{journal}{\emph{Journal of personality assessment}} \bibinfo{volume}{93}, \bibinfo{number}{1} (\bibinfo{year}{2011}), \bibinfo{pages}{84--95}.
\newblock


\bibitem[Runeson and H{\"o}st(2009)]%
        {runeson2009guidelines}
\bibfield{author}{\bibinfo{person}{Per Runeson} {and} \bibinfo{person}{Martin H{\"o}st}.} \bibinfo{year}{2009}\natexlab{}.
\newblock \showarticletitle{Guidelines for conducting and reporting case study research in software engineering}.
\newblock \bibinfo{journal}{\emph{Empirical software engineering}}  \bibinfo{volume}{14} (\bibinfo{year}{2009}), \bibinfo{pages}{131--164}.
\newblock


\bibitem[Salovey and Mayer(1990)]%
        {salovey1990emotional}
\bibfield{author}{\bibinfo{person}{Peter Salovey} {and} \bibinfo{person}{John~D Mayer}.} \bibinfo{year}{1990}\natexlab{}.
\newblock \showarticletitle{Emotional intelligence}.
\newblock \bibinfo{journal}{\emph{Imagination, cognition and personality}} \bibinfo{volume}{9}, \bibinfo{number}{3} (\bibinfo{year}{1990}), \bibinfo{pages}{185--211}.
\newblock


\bibitem[Schmelzer(2015)]%
        {schmelzer2015empathie}
\bibfield{author}{\bibinfo{person}{Daniel Schmelzer}.} \bibinfo{year}{2015}\natexlab{}.
\newblock \emph{\bibinfo{title}{Empathie in Teams: Empathieverhalten und Empathiekreisl{\"a}ufe}}.
\newblock \bibinfo{thesistype}{Ph.\,D. Dissertation}. \bibinfo{school}{M{\"u}nchen, Technische Universit{\"a}t M{\"u}nchen, Diss., 2015}.
\newblock


\bibitem[Sch{\"o}n(2017)]%
        {schon2017reflective}
\bibfield{author}{\bibinfo{person}{Donald~A Sch{\"o}n}.} \bibinfo{year}{2017}\natexlab{}.
\newblock \bibinfo{booktitle}{\emph{The reflective practitioner: How professionals think in action} (\bibinfo{edition}{1st} ed.)}.
\newblock \bibinfo{publisher}{Routledge}, \bibinfo{address}{London}.
\newblock
\urldef\tempurl%
\url{https://doi.org/10.4324/9781315237473}
\showDOI{\tempurl}


\bibitem[Schwartz and Unger(2010)]%
        {schwartz2010biculturalism}
\bibfield{author}{\bibinfo{person}{Seth~J Schwartz} {and} \bibinfo{person}{Jennifer~B Unger}.} \bibinfo{year}{2010}\natexlab{}.
\newblock \showarticletitle{Biculturalism and context: What is biculturalism, and when is it adaptive?: Commentary on Mistry and Wu}.
\newblock \bibinfo{journal}{\emph{Human development}} \bibinfo{volume}{53}, \bibinfo{number}{1} (\bibinfo{year}{2010}), \bibinfo{pages}{26}.
\newblock


\bibitem[Snow(2000)]%
        {snow2000empathy}
\bibfield{author}{\bibinfo{person}{Nancy~E. Snow}.} \bibinfo{year}{2000}\natexlab{}.
\newblock \showarticletitle{Empathy}.
\newblock \bibinfo{journal}{\emph{American Philosophical Quarterly}} \bibinfo{volume}{37}, \bibinfo{number}{1} (\bibinfo{year}{2000}), \bibinfo{pages}{65--78}.
\newblock
\showISSN{00030481}
\urldef\tempurl%
\url{http://www.jstor.org/stable/20009985}
\showURL{%
\tempurl}


\bibitem[Taleghani et~al\mbox{.}(2018)]%
        {taleghani2018barriers}
\bibfield{author}{\bibinfo{person}{Fariba Taleghani}, \bibinfo{person}{Elaheh Ashouri}, \bibinfo{person}{Mehrdad Memarzadeh}, {and} \bibinfo{person}{Mortaza Saburi}.} \bibinfo{year}{2018}\natexlab{}.
\newblock \showarticletitle{Barriers to empathy-based care: oncology nurses’ perceptions}.
\newblock \bibinfo{journal}{\emph{International Journal of Health Care Quality Assurance}} \bibinfo{volume}{31}, \bibinfo{number}{3} (\bibinfo{year}{2018}), \bibinfo{pages}{249--259}.
\newblock


\bibitem[Taylor and Bogdan(1984)]%
        {taylor1984introduction}
\bibfield{author}{\bibinfo{person}{Steven~J Taylor} {and} \bibinfo{person}{Robert~C Bogdan}.} \bibinfo{year}{1984}\natexlab{}.
\newblock \bibinfo{booktitle}{\emph{Introduction to qualitative research methods: The search for meanings}}.
\newblock \bibinfo{publisher}{Wiley}, \bibinfo{address}{New York}.
\newblock
\showISBNx{978-0-471-88947-2}


\bibitem[Taylor et~al\mbox{.}(2015)]%
        {taylor2015introduction}
\bibfield{author}{\bibinfo{person}{Steven~J Taylor}, \bibinfo{person}{Robert~C Bogdan}, {and} \bibinfo{person}{Marjorie~L. DeVault}.} \bibinfo{year}{2015}\natexlab{}.
\newblock \bibinfo{booktitle}{\emph{Introduction to Qualitative Research Methods: A Guidebook and Resource}}.
\newblock \bibinfo{publisher}{Wiley}, \bibinfo{address}{New York}.
\newblock
\showISBNx{978-1-118-76721-4}


\bibitem[Thompson et~al\mbox{.}(2019)]%
        {thompson2019empathy}
\bibfield{author}{\bibinfo{person}{Nicholas~M. Thompson}, \bibinfo{person}{Andero Uusberg}, \bibinfo{person}{James~J. Gross}, {and} \bibinfo{person}{Bhismadev Chakrabarti}.} \bibinfo{year}{2019}\natexlab{}.
\newblock \showarticletitle{Chapter 12 - Empathy and emotion regulation: An integrative account}.
\newblock In \bibinfo{booktitle}{\emph{Emotion and Cognition}}, \bibfield{editor}{\bibinfo{person}{Narayanan Srinivasan}} (Ed.). \bibinfo{series}{Progress in Brain Research}, Vol.~\bibinfo{volume}{247}. \bibinfo{publisher}{Elsevier}, \bibinfo{pages}{273--304}.
\newblock
\showISSN{0079-6123}
\urldef\tempurl%
\url{https://doi.org/10.1016/bs.pbr.2019.03.024}
\showDOI{\tempurl}


\bibitem[Wallmark et~al\mbox{.}(2018)]%
        {wallmark2018neurophysiological}
\bibfield{author}{\bibinfo{person}{Zachary Wallmark}, \bibinfo{person}{Choi Deblieck}, {and} \bibinfo{person}{Marco Iacoboni}.} \bibinfo{year}{2018}\natexlab{}.
\newblock \showarticletitle{Neurophysiological effects of trait empathy in music listening}.
\newblock \bibinfo{journal}{\emph{Frontiers in behavioral neuroscience}}  \bibinfo{volume}{12} (\bibinfo{year}{2018}), \bibinfo{pages}{66}.
\newblock
\urldef\tempurl%
\url{https://doi.org/10.3389/fnbeh.2018.00066}
\showDOI{\tempurl}


\bibitem[Wohlin et~al\mbox{.}(2012)]%
        {Wohlin2012Experimentation}
\bibfield{author}{\bibinfo{person}{Claes Wohlin}, \bibinfo{person}{Per Runeson}, \bibinfo{person}{Martin Höst}, \bibinfo{person}{Magnus~C. Ohlsson}, \bibinfo{person}{Björn Regnell}, {and} \bibinfo{person}{Anders Wesslén}.} \bibinfo{year}{2012}\natexlab{}.
\newblock \bibinfo{booktitle}{\emph{Experimentation in software engineering} (\bibinfo{edition}{1st} ed.)}. Vol.~\bibinfo{volume}{9783642290442}.
\newblock \bibinfo{publisher}{Springer}, \bibinfo{address}{Berlin, Heidelberg}.
\newblock
\urldef\tempurl%
\url{https://doi.org/10.1007/978-3-642-29044-2}
\showDOI{\tempurl}


\bibitem[Wu and Keysar(2007)]%
        {wu2007effect}
\bibfield{author}{\bibinfo{person}{Shali Wu} {and} \bibinfo{person}{Boaz Keysar}.} \bibinfo{year}{2007}\natexlab{}.
\newblock \showarticletitle{The effect of culture on perspective taking}.
\newblock \bibinfo{journal}{\emph{Psychological science}} \bibinfo{volume}{18}, \bibinfo{number}{7} (\bibinfo{year}{2007}), \bibinfo{pages}{600--606}.
\newblock


\bibitem[Xu et~al\mbox{.}(2009)]%
        {xu2009pain}
\bibfield{author}{\bibinfo{person}{Xiaojing Xu}, \bibinfo{person}{Xiangyu Zuo}, \bibinfo{person}{Xiaoying Wang}, {and} \bibinfo{person}{Shihui Han}.} \bibinfo{year}{2009}\natexlab{}.
\newblock \showarticletitle{Do You Feel My Pain? Racial Group Membership Modulates Empathic Neural Responses}.
\newblock \bibinfo{journal}{\emph{Journal of Neuroscience}} \bibinfo{volume}{29}, \bibinfo{number}{26} (\bibinfo{year}{2009}), \bibinfo{pages}{8525--8529}.
\newblock
\showISSN{0270-6474}
\urldef\tempurl%
\url{https://doi.org/10.1523/JNEUROSCI.2418-09.2009}
\showDOI{\tempurl}
\showeprint{https://www.jneurosci.org/content/29/26/8525.full.pdf}


\bibitem[Yaghoubi~Jami et~al\mbox{.}(2021)]%
        {yaghoubi2021histories}
\bibfield{author}{\bibinfo{person}{Parvaneh Yaghoubi~Jami}, \bibinfo{person}{Hyemin Han}, \bibinfo{person}{Stephen~J Thoma}, \bibinfo{person}{Behzad Mansouri}, {and} \bibinfo{person}{Rick Houser}.} \bibinfo{year}{2021}\natexlab{}.
\newblock \showarticletitle{Do histories of painful life experiences affect the expression of empathy among young adults? An electroencephalography study}.
\newblock \bibinfo{journal}{\emph{Frontiers in Psychology}}  \bibinfo{volume}{12} (\bibinfo{year}{2021}), \bibinfo{pages}{689304}.
\newblock


\bibitem[Yaghoubi~Jami et~al\mbox{.}(2023)]%
        {yaghoubi2023young}
\bibfield{author}{\bibinfo{person}{Parvaneh Yaghoubi~Jami}, \bibinfo{person}{David~Ian Walker}, {and} \bibinfo{person}{Stephen~J Thoma}.} \bibinfo{year}{2023}\natexlab{}.
\newblock \showarticletitle{Young adults’ empathic responses to others in psychological pain as compared to physical pain: does prior experience of pain matter?}
\newblock \bibinfo{journal}{\emph{Current Psychology}} \bibinfo{volume}{42}, \bibinfo{number}{8} (\bibinfo{year}{2023}), \bibinfo{pages}{6194--6215}.
\newblock


\bibitem[Yu and Kirk(2009)]%
        {yu2009evaluation}
\bibfield{author}{\bibinfo{person}{Juping Yu} {and} \bibinfo{person}{Maggie Kirk}.} \bibinfo{year}{2009}\natexlab{}.
\newblock \showarticletitle{Evaluation of empathy measurement tools in nursing: systematic review}.
\newblock \bibinfo{journal}{\emph{Journal of advanced nursing}} \bibinfo{volume}{65}, \bibinfo{number}{9} (\bibinfo{year}{2009}), \bibinfo{pages}{1790--1806}.
\newblock


\end{thebibliography}

\appendix
\section{Memos} \label{SEC:MemosAppendix}
    \begin{boxA}
    \footnotesize
    \textbf{Memo on ``Developer Technical Support''}
    
    When we asked developers for examples of empathising with users, some stated that they modified the questions based on the scenarios and personalised the content for users during the usability sessions especially when users seemed confused [P1]. They stated that it was an act of consideration towards each individual user. Also, when asked about how well developers think users understood the shared technical details developers stated that they implicitly tried to talk in layman's terms when dealing with the users [P3]. Developers stated that they are confident that the users sufficiently understood the technical details as clearly demonstrated by users' accurate actions during usability session tasks. This was confirmed by the responses of the users. All the users stated that they were able to clearly understand and follow developer instructions. This suggests that modifying or personalising content for easier user understanding is one of the first things that comes to their mind when thinking of empathy. When we inquired about user’s understanding of technical details, developers stated that they didn’t want to overload users with technical details that they couldn’t understand. They also said that they explicitly tried to phrase questions as clearly as possible for first time as they get quite nervous while hosting the usability sessions [P1, P6]. This demonstrates that the developers tried to support the users by giving the best possible instructions. 
    \end{boxA}
    

\section{Interview Guides} \label{SEC:Interview GuidesAppendix}
\subsection{Interview Guide of Software Developers} 
\textbf{General Information:} 
\begin{enumerate}
    \item Do you study a double degree?
    \item (If Yes) What are your majors?
    \item (If following a non-SE Degree) I want to learn a bit more about the non-SE degree that you follow. Do you have close contact with customers outside the university, for example during your projects and internships?
    \item I am going to ask you to rate your affinity to technology vs people. If we consider a scale of 4, where 0 is very human-centric (more affinity to people), 1 is somewhat human-centric, 2 somewhat technology-centric and 3 fully technology-centric, how would you rate yourself? 
\end{enumerate}

\textbf{Related to the experience of interacting with end users:}

In this study, empathy refers to understanding a person from his or her frame of reference rather than one’s own, or vicariously experiencing that person’s feelings, perceptions, and thoughts. 
\begin{enumerate}
\setcounter{enumi}{4}
    \item Based on this definition, how well do you think you were able to empathise with the end users? Why do you think so? Can you share an example of a time when you were able to empathise with the end users?
    \item  Were there any instances where it was difficult for you to empathise with the end users? Can you share an example?
    \item Thinking back now, was there a time when you could have empathised better with the end users?
    \item If you could do it again, is there anything you would like to change with respect to engaging with the end users / empathising with them? Why?
    \item Do you recognise any special reasons as to why you were able to empathise with end users or why you were unable to empathise with the end users? Can you share an example?
    \item  When you explain certain technical aspects or limitations in the application, do you think end users were able to fully understand what you explained? Why do you think so?
    \item How did you feel when you were conducting the usability sessions?
    \item Finally, is there anything else you would like to share? Any further feedback for us on your experience on this project?

\end{enumerate}

\subsection{Interview Guide of End Users}
\textbf{General Information:} 
\begin{enumerate}
    \item Have you ever worked with software developers before? Or do you have any close contact with software developers?
    \item I am going to ask you to rate your affinity to technology vs people. If we consider a scale of 4, where 0 is very human-centric (more affinity to people), 1 is somewhat human-centric, 2 somewhat technology-centric and 3 fully technology-centric, how would you rate yourself? 
\end{enumerate}

\textbf{Related to the experience of interacting with end users:}

In this study, empathy refers to understanding a person from his or her frame of reference rather than one’s own, or vicariously experiencing that person’s feelings, perceptions, and thoughts. 
\begin{enumerate}
\setcounter{enumi}{2}
    \item Based on this definition, how well do you think you were able to empathise with the developers? Why do you think so? Can you share an example of a time when you were able to empathise with the developers?
    \item  Were there any instances where it was difficult for you to empathise with the developers? Can you share an example?
    \item Thinking back now, was there a time when you could have empathised better with the developers?
    \item If you could do it again, is there anything you would like to change with respect to engaging with the developers / empathising with them? Why?
    \item Do you recognise any special reasons as to why you were able to empathise with developers or why you were unable to empathise with them? Can you share an example?
    \item  When the developers explained certain technical aspects or limitations in the application, do you think you were able to fully understand what they explained? Why do you think so?
    \item How did you feel when you were participating in the usability sessions?
    \item Finally, is there anything else you would like to share regarding the study? Any further feedback for us on your experience on this project?

\end{enumerate}


\end{document}